\documentclass[11pt]{article}
\usepackage{a4wide}
\usepackage{graphicx}
\newcommand{\osigma}{\overline{\sigma}}
\newcommand{\oc}{\overline{c}}
\newcommand{\text}{\rm}
\newcommand{\odelta}{\overline{\delta}}
\newcommand{\os}{\overline{s}}

\newcommand{\sect}[1]{ \section{#1} \setcounter{equation}{0} }

\begin{document}

\title{ \textbf{More on ghost condensation in Yang-Mills theory: BCS versus Overhauser
effect and the breakdown of the Nakanishi-Ojima annex $SL(2,R)$
symmetry}}

\author{D. Dudal\thanks{%
Research Assistant of The Fund For Scientific Research-Flanders,
Belgium.}
\ and H. Verschelde\thanks{%
david.dudal@rug.ac.be, henri.verschelde@rug.ac.be} \\
{\small {\textit{Ghent University }}}\\
{\small {\textit{Department of Mathematical Physics and Astronomy,
Krijgslaan 281-S9, }}}\\
{\small {\textit{B-9000 Gent, Belgium}}} \and V.E.R. Lemes, M.S.
Sarandy
and S.P. Sorella\thanks{%
vitor@dft.if.uerj.br, sarandy@dft.if.uerj.br, sorella@uerj.br} \\
{\small {\textit{UERJ - Universidade do Estado do Rio de Janeiro,}}} \\
{\small {\textit{\ Rua S\~{a}o Francisco Xavier 524, 20550-013
Maracan\~{a},
}}} {\small {\textit{Rio de Janeiro, Brazil}}} \and M. Picariello and A. Vicini\thanks{%
marco.picariello@mi.infn.it, alessandro.vicini@mi.infn.it } \\
{\small {\textit{Universit\'{a} degli Studi di Milano, via Celoria
16,
I-20133, Milano, Italy }}}\\
{\small {\textit{and INFN\ Milano, Italy}}}
\and J.A. Gracey\thanks{jag@amtp.liv.ac.uk } \\
{\small {\textit{Theoretical Physics Division}}} \\
{\small {\textit{Department of Mathematical Sciences}}} \\
{\small {\textit{University of Liverpool }}} \\
{\small {\textit{P.O. Box 147, Liverpool, L69 3BX, United
Kingdom}}}}  \maketitle

\vspace{-14cm} \hfill LTH--579 \vspace{14cm}

\begin{abstract}
We analyze the ghost condensates $\left\langle
f^{abc}c^{b}c^{c}\right\rangle$, $\left\langle
f^{abc}\oc^{b}\oc^{c}\right\rangle$ and $\left\langle
f^{abc}\oc^{b}c^{c}\right\rangle$ in Yang-Mills theory in the
Curci-Ferrari gauge.  By combining the local composite operator
formalism with the algebraic renormalization technique, we are
able to give a simultaneous discussion of $\left\langle
f^{abc}c^{b}c^{c}\right\rangle$,$\left\langle
f^{abc}\oc^{b}\oc^{c}\right\rangle$ and $\left\langle
f^{abc}\oc^{b}c^{c}\right\rangle$, which can be seen as playing
the role of the BCS, respectively Overhauser effect in ordinary
superconductivity. The Curci-Ferrari gauge exhibits a global
continuous symmetry generated by the Nakanishi-Ojima ($NO$)
algebra. This algebra includes, next to the \mbox{(anti-)BRST}
transformation, a $SL(2,R)$ subalgebra. We discuss the dynamical
symmetry breaking of the $NO$ algebra through these ghost
condensates. Particular attention is paid to the Landau gauge, a
special case of the Curci-Ferrari gauge.
\end{abstract}

\makeatother

\sect{Introduction} Vacuum condensates play an important role in
quantum field theory. They can be used to parametrize some
non-perturbative effects. If one wants to attach a physical
meaning to a certain condensate in case of a gauge theory, it
should evidently be gauge invariant. Two well known examples in
the context of QCD are the gluon condensate $\left\langle
F_{\mu\nu}^{2}\right\rangle$ and the
quark condensate $\left\langle \overline{q}q\right\rangle$. \\\\
Recently, there was a growing interest for a mass dimension 2
condensate in (quarkless) QCD in the Landau gauge, see e.g.
\cite{Gubarev:2000eu,Gubarev:2000nz,Boucaud:2000ey,
Boucaud:2001st,Boucaud:2002nc,Verschelde:2001ia}. Unfortunately,
no local gauge invariant operator with mass dimension 2 exists.
However, a non-local gauge invariant dimension 2 operator can be
constructed by minimizing $A^{2}$ along each gauge orbit, namely
$A^{2}_{\tiny{\textrm{min}}}\equiv (VT)^{-1}\min_{U}\int
d^{4}x\left(A_{\mu}^{U}\right)^{2}$ with $VT$ the space time
volume and $U$ a generic $SU(N)$ transformation. This operator is
related to the Gribov region as well as the so-called fundamental
modular region (FMR), which is the set of absolute minima of $\int
d^{4}x\left(A_{\mu}^{U}\right)^{2}$
\cite{semenov,Zwanziger:tn,Stodolsky:2002st}. In particular, in
the Landau gauge $\partial_{\mu}A^{\mu}=0$, it turned out that
$A^{2}_{\tiny{\textrm{min}}}$ reduces to the local operator
$A^{2}$. This gives a meaning to the condensate $\left\langle
A^{2}\right\rangle$. In \cite{Verschelde:2001ia}, an effective
action was constructed in the weak coupling for the $\left\langle
A^{2}\right\rangle$ condensate by means of the local composite
operator technique (LCO) and it was shown that $\left\langle
A^{2}\right\rangle\neq0$ is dynamically favoured since it lowers
the vacuum energy. Due to
this condensate, the gluons achieved a mass.\\\\
In this article, we will discuss other condensates of mass
dimension 2 \cite{Dudal:2003pw}, namely pure ghost condensates of
the type $\left\langle f^{abc}c^{b}c^{c}\right\rangle$,
$\left\langle f^{abc}\oc^{b}\oc^{c}\right\rangle$ and
$\left\langle f^{abc}\oc^{b}c^{c}\right\rangle$. Historically,
these condensates came to attention in
\cite{Schaden:1999ew,Schaden:2000fv,Schaden:2001xu,Kondo:2000ey}
in the context of $SU(2)$ Yang-Mills theory in the Maximal Abelian
gauge. This is a partial non-linear gauge fixing which requires
the introduction of a four ghost interaction term for consistency.
A decomposition, by means of a Hubbard-Stratonovich auxiliary
field, similar to the one of the 4-fermion interaction of the
Gross-Neveu model \cite{Gross:jv}, allowed to construct a 1-loop
effective potential, leading to a non-trivial minimum for the
ghost condensate corresponding to $\left\langle
f^{abc}\oc^{b}c^{c}\right\rangle$. It was recognized in
\cite{Schaden:1999ew,Schaden:2000fv,Schaden:2001xu} that this
condensate signals the breakdown of a global $SL(2,R)$ symmetry of
the $SU(2)$ Maximal Abelian gauge model. The ghost condensate was
used to find a mass for the off-diagonal gluons, and thereby a
certain evidence for the Abelian dominance was established
\cite{Kondo:2000ey}. It has been shown since then that the ghost
condensate gives in fact a tachyonic mass \cite{Dudal:2002xe}.
\\\\It is worth mentioning that a simple decomposition of the
4-fermion interaction might cause troubles with the
renormalizability beyond the 1-loop order. For instance, in the
case of the Gross-Neveu model, this procedure requires the
introduction of ad hoc counterterms to maintain finiteness
\cite{Verschelde:jx,Luperini:1991sv}. A similar problem can be
expected with the 4-ghost interaction. The
LCO procedure gave an outcome to this problem \cite{Verschelde:jx}.\\\\
Another issue that deserves clarification is the fact that with a
different decomposition, different ghost condensates appear
\cite{Lemes:2002ey}, corresponding to the Faddeev-Popov charged
condensates $\left\langle f^{abc}c^{b}c^{c}\right\rangle$ and
$\left\langle f^{abc}\oc^{b}\oc^{c}\right\rangle$. The existence
of several channels for the ghost condensation has a nice analogy
in the theory of superconductivity, known as the BCS versus
Overhauser effect. The BCS channel corresponds to the charged
particle-particle and hole-hole pairing
\cite{Bardeen:1957kj,Bardeen:1957mv}, while the Overhauser channel
to the particle-hole pairing \cite{overhauser,Park:1999bz}. In the
present case, the Faddeev-Popov charged condensates $\left\langle
f^{abc}\oc^{b}\oc^{c}\right\rangle$ and $\left\langle
f^{abc}c^{b}c^{c}\right\rangle$ would correspond to the BCS
channel, while $\left\langle f^{abc}\oc^{b}c^{c}\right\rangle$ to
the Overhauser channel. The question is whether one of these
effects would be favoured. A simultaneous discussion of both
effects is necessary to find out if one vacuum is more stable than
the other.
\\\\It is appealing that
by now the ghost condensates have been observed also in a class of
non-linear generalized covariant gauges
\cite{Delbourgo:1981cm,Baulieu:sb}, the so-called Curci-Ferrari
gauges\footnote{Referring to the massive Curci-Ferrari model that
has the same gauge fixing terms \cite{Curci:bt,Curci:1976ar}.},
again by the decomposition of a 4-ghost interaction
\cite{Lemes:2002jv}. The Curci-Ferrari gauge has the Landau gauge
as a special case. Although the Landau gauge lacks a 4-ghost
interaction, it has been shown that the ghost condensation also
takes place in this gauge \cite{Lemes:2002rc}. Evidently, this was
not possible by the decomposition of a 4-point interaction.
However, the combination of the LCO method
\cite{Verschelde:2001ia,Knecht:2001cc} with the algebraic
renormalization formalism \cite{book,Barnich:2000zw} allowed for a
clean treatment
of the ghost condensation in the Landau gauge.\\\\
It seems thus that the ghost condensation takes place in a variety
of gauges: the Landau gauge, the Curci-Ferrari gauge and the
Maximal Abelian gauge. It is known that the Landau gauge and
Curci-Ferrari gauge exhibit a global continuous symmetry,
generated by the so-called Nakanishi-Ojima algebra
\cite{Nakanishi:1980dc,Delduc:1989uc,Sawayanagi:zw,Nishijima:1984gj,Nishijima:qt,
book2}. This algebra contains, next to the BRST and anti-BRST
transformations, a $SL(2,R)$ subalgebra generated by the
Faddeev-Popov ghost number and 2 other operators, $\delta$ and
$\odelta$. Moreover, $\delta$ and $\odelta$ mutually transform the
ghost operators $f^{abc}c^{b}c^{c}$, $f^{abc}\oc^{b}\oc^{c}$ and
$f^{abc}\oc^{b}c^{c}$ into each other. It is then apparent that
the ghost condensation can appear in several channels like the BCS
and Overhauser channel, and that a non-vanishing vacuum
expectation value for the ghost operators indicates a breakdown of
this $SL(2,R)$ symmetry.
\\\\Recently, it has been shown that the same\footnote{The
$SL(2,R)$ symmetry discussed in
\cite{Schaden:1999ew,Schaden:2000fv,Schaden:2001xu,Sawayanagi:zw}
is only acting non-trivially on the off-diagonal fields.} $NO$
invariance of the Landau and \mbox{Curci-Ferrari} gauge can be
maintained in the Maximal Abelian gauge for any value of $N$
\cite{Dudal:2002ye}. Apparently, an intimate connection exists
between the $NO$ symmetry and the appearance of the ghost
condensates, since all gauges where
the ghost condensates has been proven to occur, have the global $NO$ invariance.\\\\
The aim of this article is to provide an answer to the
aforementioned issues. We will discuss the Curci-Ferrari gauge.
For explicit calculations, we will restrict ourselves to the
Landau gauge for $SU(2)$. The presented general arguments are
however neither depending on the choice of the gauge parameter,
nor on the value of $N$. The paper is organized as follows. In
section 2, we show that it is possible to introduce a set of
external sources for the ghost operators, according to the LCO
method, and this without spoiling the $NO$ invariance. Employing
the algebraic renormalization technique
\cite{book,Barnich:2000zw}, it can then be checked that the
proposed action can be renormalized. In section 3, the effective
potential for the ghost condensates is evaluated. By contruction,
this effective potential, incorporating the BCS as well as the
Overhauser channel, is finite up to \emph{any order} and obeys a
homogeneous renormalization group equation. Next, in section 4, we
pay attention to the dynamical symmetry breaking of the $NO$
algebra due to the ghost condensates. Because of the $SL(2,R)$
invariance of the presented framework, it becomes clear that a
whole class of equivalent, non-trivial vacua exist. The Overhauser
and the BCS vacuum are important special cases. Notice that a
nonvanishing condensate $\left\langle
f^{abc}c^{b}c^{c}\right\rangle\neq0$ could seem to pose a problem
for the Faddeev-Popov ghost number symmetry and for the BRST
symmetry, two basic properties of a quantized gauge theory.
However, we shall be able to show that one can define \emph{a}
nilpotent BRST and \emph{a} Faddeev-Popov symmetry in any possible
ghost condensed vacuum. The existence of the $NO$ symmetry plays a
key role in this. Since the ghost condensates carry a color index,
we also spend some words on the global $SU(N)$ color symmetry.
Here, we can provide an argument that, thanks to the existence of
the condensate $\left\langle A^{2}\right\rangle$ and of its
generalization $\left\langle \frac{1}{2}A^{2}+\alpha\oc
c\right\rangle$ in the Curci-Ferrari gauge \cite{Dudal:2003gu},
the breaking of the color symmetry, induced by the ghost
condensates, should be located in the unphysical part of the
Hilbert space. Furthermore, we argue why no physical Goldstone
particles should appear by means of the quartet mechanism
\cite{Kugo:gm}. Section 5 handles the generalization of the
results to the case with quarks included. In section 6, we give an
outline of future research where the gluon and ghost condensates
can play a role. We end with conclusions in section 7. Technical
details are collected in the Appendices A and B.

\sect{The set of external sources for both BCS and Overhauser
channel}
\subsection{Introduction of the LCO sources}
For a thorough introduction to the local composite operator (LCO)
formalism and to the algebraic renormalization technique, the
reader is
referred to \cite{Verschelde:2001ia,Knecht:2001cc}, respectively \cite{book}.\\\\
According to the LCO method, the first step in the analysis of the
ghost condensation in both channels is the introduction of a
suitable system of external sources. Generalizing the construction
done in the pure BCS case \cite{Lemes:2002rc}, it turns out that
the simultaneous presence of both channels is achieved by
considering the following BRST invariant external action
\begin{eqnarray}
S_{LCO} &=&s\int d^{4}x\left( L^{a}c^{a}+\lambda ^{a}\left(
b^{a}-gf^{abc}\overline{c}^{b}c^{c}\right) +\zeta \eta
^{a}L^{a}-\frac{1}{2}\eta
^{a}gf^{abc}\overline{c}^{b}\overline{c}^{c}+
\frac{1}{2}\rho\lambda
^{a}\omega ^{a}-\omega ^{a}\overline{c}^{a}\;\right)  \nonumber \\
&=&\int d^{4}x\left(
\frac{1}{2}L^{a}gf^{abc}c^{b}c^{c}-\frac{1}{2}\tau
^{a}gf^{abc}\overline{c}^{b}\overline{c}^{c}+\eta ^{a}gf^{abc}b^{b}\overline{%
c}^{c}+\zeta \tau ^{a}L^{a}\right.  \nonumber \\
 &-&\left.\omega ^{a}gf^{abc}\overline{c}^{b}c^{c}+\;\lambda
^{a}gf^{abc}b^{b}c^{c}-\frac{1}{2}\lambda ^{a}g^{2}f^{abc}\overline{c}^{b}%
f^{cmn}c^{m}c^{n}+ \frac{1}{2}\rho\omega ^{a}\omega ^{a}\right)
\nonumber \\
&&  \label{slco}
\end{eqnarray}
The BRST transformation $s$ is defined for the fields $A_{\mu
}^{a}$, $c^{a}$, $\overline{c}^{a}$, $b^{a}$ as
\begin{eqnarray}
sA_{\mu }^{a} &=&-D_{\mu }^{ab}c^{b}  \nonumber \\
sc^{a} &=&\frac{g}{2}f^{abc}c^{b}c^{c}  \nonumber \\
s\overline{c}^{a} &=&b^{a}  \nonumber \\
sb^{a} &=&0  \label{s}
\end{eqnarray}
with
\begin{equation}
D_{\mu }^{ab}= \partial _{\mu }\delta ^{ab}+gf^{acb}A_{\mu }^{c}
\label{dudal0}
\end{equation}
the adjoint covariant derivative. \\\\The external sources
$L^{a}$, $\tau ^{a}$, $\lambda ^{a}$, $\omega ^{a} $, $\eta ^{a}$
transform as
\begin{eqnarray}
s\eta ^{a} &=&\tau ^{a}\;,\;\;\;\;s\tau ^{a}=\;0\;,  \label{lcos} \\
s\lambda ^{a} &=&\omega ^{a}\;,\;\;\;s\omega ^{a}=\;0\;,  \nonumber \\
sL^{a} &=&0  \nonumber
\end{eqnarray}
\begin{equation}
\begin{tabular}{|l|c|c|c|c|c|}
\hline
& $L^{a}$ & $\eta ^{a}$ & $\tau ^{a}$ & $\lambda ^{a}$ & $\omega ^{a}$ \\
\hline $\mathrm{Dimension}$ & $2$ & $1$ & $2$ & $1$ & $2$ \\
\hline $\mathrm{Gh.\;Number}$ & $-2$ & $1$ & $2$ & $-1$ & $0$ \\
\hline
\end{tabular}
\label{table1}
\end{equation}
>From expression (\ref{slco}) one sees that the sources $L^{a}$,
$\tau ^{a}$
couple to the ghost operators $gf^{abc}c^{b}c^{c}$, $gf^{abc}\overline{c}^{b}%
\overline{c}^{c}$ of the BCS channel, while $\omega ^{a}$ accounts
for the Overhauser channel $gf^{abc}\overline{c}^{b}c^{c}$. As far
as the BRST invariance is the only invariance required for the
external action (\ref {slco}), the LCO parameters $\zeta $ and
$\rho $ are independent. However, it is known that both the Landau
and the Curci-Ferrari gauge display a larger set of symmetries,
giving rise to the $NO$ algebra
\cite{Delbourgo:1981cm,Baulieu:sb,Nakanishi:1980dc,Delduc:1989uc,Sawayanagi:zw,Nishijima:1984gj,Nishijima:qt,book2,Dudal:2002ye}.
It is worth remarking that the whole $NO$ algebra can be extended
also in the presence of the external action $S_{LCO}$, provided
that the two parameters $\zeta $ and $\rho $ obey the relationship
\begin{equation}
\rho =2\zeta   \label{rel}
\end{equation}
In other words, the requirement of invariance of $S_{LCO}$ under
the whole $NO$ algebra allows for a unique parameter in expression
(\ref{slco}). In order to introduce the generators of the $NO$
algebra, let us begin with the anti-BRST transformation $\os$
\begin{eqnarray}
\overline{s}A_{\mu }^{a} &=&-D_{\mu }^{ab}\overline{c}^{b}  \nonumber \\
\overline{s}c^{a} &=&-b^{a}+gf^{abc}c^{b}\overline{c}^{c}  \nonumber \\
\overline{s}\overline{c}^{a} &=&\frac{g}{2}f^{abc}\overline{c}^{b}\overline{c%
}^{c}  \nonumber \\
\overline{s}b^{a} &=&-gf^{abc}b^{b}\overline{c}^{c}  \label{as}
\end{eqnarray}
Extending $\overline{s}$ to the external LCO sources as
\begin{eqnarray}
\overline{s}\eta ^{a} &=&-\omega ^{a}\;,\;\;\;\;\overline{s}\tau
^{a}=\;0
\label{aslco} \\
\overline{s}\lambda ^{a} &=&L^{a}\;,\;\;\;\;\;\;\overline{s}\omega
^{a}=\;0  \nonumber \\
\overline{s}L^{a} &=&0  \nonumber
\end{eqnarray}
one easily verifies that
\begin{equation}
\{s,\os\}=s\overline{s}+\overline{s}s=0  \label{sas}
\end{equation}
Furthermore, the requirement of invariance of $S_{LCO}$ under $%
\overline{s}$ fixes the parameter $\rho =2\zeta $, namely
\begin{equation}
\overline{s}S_{LCO}=0\;\Rightarrow \;\rho =2\zeta \label{asr}
\end{equation}
This is best seen by observing that, when $\rho =2\zeta $, the
whole action $S_{LCO}$ can be written as
\begin{equation}\label{l1}
S_{LCO}=s\overline{s}\int d^{4}x\left( \lambda ^{a}c^{a}+\zeta
\lambda ^{a}\eta ^{a}+\eta ^{a}\overline{c}^{a}\right)
\end{equation}
Concerning now the other generators $\delta $ and
$\overline{\delta }$ of the $NO$ algebra, they can be introduced
as follows
\begin{eqnarray}
\delta \overline{c}^{a} &=&c^{a}  \nonumber \\
\delta b^{a} &=&\frac{g}{2}f^{abc}c^{b}c^{c}  \nonumber \\
\delta A_{\mu }^{a} &=&\delta c^{a}=0  \nonumber \\
\delta L^{a} &=&2\omega ^{a}  \nonumber \\
\delta \omega ^{a} &=&-\tau ^{a}  \nonumber \\
\delta \lambda ^{a} &=&-\eta ^{a}  \nonumber \\
\delta \tau ^{a} &=&\delta \eta ^{a}=0  \label{d}
\end{eqnarray}
and
\begin{eqnarray}
\overline{\delta }c^{a} &=&\overline{c}^{a}  \nonumber \\
\overline{\delta }b^{a} &=&\frac{g}{2}f^{abc}\overline{c}^{b}\overline{c}%
^{c}  \nonumber \\
\overline{\delta }A_{\mu }^{a} &=&\overline{\delta
}\overline{c}^{a}=0
\nonumber \\
\overline{\delta }\omega ^{a} &=&L^{a}  \nonumber \\
\overline{\delta }\tau ^{a} &=&-2\omega ^{a}  \nonumber \\
\overline{\delta }\eta ^{a} &=&-\lambda ^{a}  \nonumber \\
\overline{\delta }L^{a} &=&\overline{\delta }\lambda ^{a}=0
\label{ad}
\end{eqnarray}
It holds that
\begin{equation}
\delta S_{LCO}=\overline{\delta }S_{LCO}=0 \label{dadi}
\end{equation}
The operators $s$, $\overline{s}$, $\delta $, $\overline{\delta }$
and the Faddeev-Popov ghost number operator $\delta _{FP}$ give
rise to the $NO$ algebra
\begin{eqnarray}
s^{2} &=&0\,,\,\,\,\,\,\,\,\,\,\,\,\overline{s}^{2}=0  \nonumber \\
\,\{s,\overline{s}\} &=&0\,,\,\,\,\,\,\,\,\,\,\,[\delta ,\overline{\delta }%
]=\delta _{\mathrm{FP}}\,,  \nonumber \\
\lbrack \delta ,\delta _{\mathrm{FP}}] &=&-2\delta ,\,\,\,[\overline{\delta }%
,\delta _{\mathrm{FP}}]=2\overline{\delta }  \nonumber \\
\lbrack s,\delta _{\mathrm{FP}}] &=&-s\,,\,\,\,\,\,\,[\overline{s},\delta _{%
\mathrm{FP}}]=\overline{s}\,,  \nonumber \\
\lbrack s,\delta ] &=&0\,\,,\,\,\,\,\,\,\,\,[\overline{s},\overline{\delta }%
]=0\,,  \nonumber \\
\lbrack s,\overline{\delta }] &=&-\overline{s}\,\,,\,\,\,\,\,[\overline{s}%
,\delta ]=-s  \label{no}
\end{eqnarray}
In
particular, $\delta _{FP}$, $\delta $, $\overline{\delta }$ generate a $%
SL(2,R)$ subalgebra. We remark that the $NO$ algebra can be
established as an exact invariance of $S_{LCO}$ only when both
channels are present. It is easy to verify indeed that setting to
zero the external sources corresponding to one channel will imply
the loss of the $NO$ algebra. This implies that a complete
discussion of the ghost condensates needs sources for the BCS as
well as for the Overhauser channel.\\\\ Let us also give, for
further use, the expressions of the gauge fixed action in the
presence of the LCO external sources for the Curci-Ferrari gauge.
\begin{eqnarray}
S &=&S_{YM}+S_{GF+FP}+S_{LCO}  \nonumber \\
&=&-\frac{1}{4}\int d^{4}xF_{\mu \nu }^{a}F^{a\mu \nu
}+s\overline{s}\int d^{4}x\left( \frac{1}{2}A_{\mu }^{a}A^{a\mu
}+\lambda ^{a}c^{a}+\zeta
\lambda ^{a}\eta ^{a}+\eta ^{a}\overline{c}^{a}-\frac{\alpha }{2}c^{a}%
\overline{c}^{a}\right)   \nonumber \\
&&  \label{scf}
\end{eqnarray}
with
\begin{eqnarray}
S_{GF+FP} &=&\int d^{4}x\left( b^{a}\partial _{\mu }A^{a\mu }+\frac{\alpha }{%
2}b^{a}b^{a}+\overline{c}^{a}\partial ^{\mu }D_{\mu
}^{ab}c^{b}-\frac{\alpha }{2}gf^{abc}b^{a}\overline{c}^{b}c^{c}
-\frac{\alpha }{%
8}g^{2}f^{abc}f^{cde}\overline{c}^{a}\overline{c}^{b}c^{d}c^{e}\right)\nonumber\\
\label{sfpcf}
\end{eqnarray}
The renormalizability of the action (\ref{scf}) is discussed in the \mbox{Appendix A}.\\\\
The Curci-Ferrari gauge has the Landau gauge, $\alpha=0$, as
interesting special case, see for example \cite{Dudal:2003gu}.
One sees that the difference between the two actions is due to the term $%
\alpha c^{a}\overline{c}^{a}$, which gives rise to a quartic ghost
self interaction absent in the Landau gauge. The whole set of $NO$
invariances can be translated into functional identities which
ensures the renormalizability of the model. In particular,
concerning the counterterm
contributions $\delta_{L}L^{a}gf^{abc}c^{b}c^{c}$, $\delta_{\tau}\tau^{a}gf^{abc}\overline{c}^{b}\overline{c}%
^{c}$ and
$\delta_{\omega}\omega^{a}gf^{abc}\overline{c}^{b}c^{c}\;$, it is
shown in the \mbox{Appendix A} that
\begin{equation}\label{gh}
\delta_{L}=\delta_{\tau}=\delta_{\omega}\equiv\delta_{2}
\end{equation}
Consequently, the operators $gf^{abc}c^{b}c^{c}$, $gf^{abc}%
\overline{c}^{b}\overline{c}^{c}$ and
$gf^{abc}\overline{c}^{b}c^{c}$ turn out to have the same
anomalous dimension for any $\alpha$. As expected, this result is
a consequence of the presence of the $NO$ symmetry. Moreover, in
the Landau gauge, $\delta_{2}\equiv 0$ due to the
nonrenormalization properties of the Landau gauge \cite{book}. In
\cite{Dudal:2002pq}, one can find an explicit proof that
$\delta_{2}=0$.

\subsection{A note on the choice of the hermiticity properties of the Faddeev-Popov ghosts}
With our choice of ghosts $c$, respectively anti-ghosts
$\overline{c}$, the hermiticity assignment
\begin{eqnarray}\label{l3}
  c^{\dag} &=& c \nonumber\\
  \overline{c}^{\dag} &=& -\overline{c}
\end{eqnarray}
is obeyed. This implies that $c$ and $\overline{c}$ are
independent degrees of freedom and by a redefinition
$i\overline{c}=\overline{c}'$, we have real (anti-)ghost fields
$c$ and $\overline{c}'$. Another assignment that is used
sometimes, reads
\begin{eqnarray}\label{l4}
  c^{\dag} &=& \overline{c}
\end{eqnarray}
As it was explored in e.g. \cite{book2,Kugo:gm}, the former
assignment is the correct one for a generic gauge. However, based
on the additional ghost-anti-ghost symmetry in the Landau gauge,
both formulations are equivalent. Moreover, this equivalence,
which is related to the existence of the $SL(2,R)$ symmetry, can
be maintained if the Landau gauge is generalized to the
Curci-Ferrari gauge \cite{Alkofer:2000wg}. Since we are discussing
the existence of $\left\langle
f^{abc}\overline{c}^{b}c^{c}\right\rangle$, $\left\langle
f^{abc}c^{b}c^{c}\right\rangle$ and $\left\langle
f^{abc}\overline{c}^{b}\overline{c}^{c}\right\rangle$, which break
the $SL(2,R)$ symmetry, the equivalence between the real
formulation (\ref{l3}) and the complex one (\ref{l4}) might be
altered. For example, if $\left\langle
f^{abc}c^{b}c^{c}\right\rangle\neq\left\langle
f^{abc}\overline{c}^{b}\overline{c}^{c}\right\rangle$, the
ghost-anti-ghost symmetry is lost, as well as the usual ghost
number symmetry. Throughout this article, we will use the
prescription (\ref{l3}). We will return to the issue of the ghost
number symmetry later in this article.

\sect{Effective potential for the ghost condensates}
\subsection{General considerations}
Let us proceed with the construction of the effective potential
for the ghost condensates in the Curci-Ferrari gauge. To decide
which channel is favoured, we have to consider the 2 channels at
once. We shall also treat the two LCO parameters $\rho $ and
$\zeta $ for the moment as being independent and verify the
relationship (\ref{slco}). Setting to zero the external sources
$\eta $ and $\lambda $, we start from the action
\begin{eqnarray}
S &=&S_{YM}+S_{GF+FP}+\int d^{4}x\left[ -\omega ^{a}gf^{abc}\overline{c}%
^{b}c^{c}+\frac{1}{2}\rho \omega ^{a}\omega ^{a}\right.  \nonumber
\label{d1} \\
&+&\left. \frac{1}{2}L^{a}gf^{abc}c^{b}c^{c}-\frac{1}{2}\tau ^{a}gf^{abc}%
\overline{c}^{b}\overline{c}^{c}+\zeta \tau ^{a}L^{a}\right]
\end{eqnarray}
Following \cite{Verschelde:2001ia,Knecht:2001cc}, the divergences
proportional to $L\tau $ are cancelled by
the counterterm $\delta \zeta \tau L$, and the divergences proportional to $%
\omega ^{2}$ are cancelled by the counterterm $\frac{\delta \rho
}{2}\omega ^{2}$. Considering the bare Lagrangian associated to
(\ref{d1}), we have
\begin{eqnarray}
  c_{b} &=& \sqrt{Z_{c}}c \hspace{1cm}\oc_{b} = \sqrt{Z_{c}}\oc \\
  A_{b} &=& \sqrt{Z_{A}}A \\
  g_{b} &=& \mu^{\varepsilon/2}Z_{g}g \\
  L_{b} &=& \mu^{-\varepsilon/2}\frac{Z_{2}}{Z_{g}Z_{c}}L
  \hspace{1cm}\tau_{b} =\mu^{-\varepsilon/2}\frac{Z_{2}}{Z_{g}Z_{c}}\tau
  \hspace{1cm}\omega_{b}= \mu^{-\varepsilon/2}\frac{Z_{2}}{Z_{g}Z_{c}}\omega
\end{eqnarray}
where $Z_{2}=1+\delta_{2}$ (see (\ref{gh})).\\ Furthermore,
\begin{eqnarray}
\zeta _{b}\tau _{b}^{a}L_{b}^{a} &=&\mu ^{-\varepsilon}\left(
\zeta +\delta \zeta \right) \tau ^{a}L^{a}  \label{d2} \\
\frac{1}{2}\rho _{b}\omega _{b}^{a}\omega _{b}^{a} &=&\frac{1}{2}\mu^{-%
\varepsilon}\left( \rho +\delta \rho \right) \omega ^{a}\omega
^{a}  \label{d2bis}
\end{eqnarray}
where it is understood that we are working with dimensional
regularization in $d=4-\varepsilon$ dimensions. The above
equations allow to derive the renormalization group equation of
$\zeta $ and $\rho$
\begin{eqnarray}
\mu \frac{d\zeta }{d\mu } &=&2\gamma (g^{2})\zeta +\delta _{\zeta
}(g^{2})
\label{d3bis} \\
\mu \frac{d\rho }{d\mu } &=&2\gamma (g^{2})\rho +\delta _{\rho
}(g^{2}) \label{d3}
\end{eqnarray}
where $\gamma (g^{2})$ denotes the anomalous dimension of the ghost operators $gf^{abc}\overline{c}^{b}%
c^{c}$, $gf^{abc}c^{b}c^{c}$ and $gf^{abc}\overline{c}^{b}%
\overline{c}^{c}$, given by
\begin{equation}\label{running}
    \gamma(g^{2})=\mu\frac{d}{d\mu}\ln\frac{Z_{2}}{Z_{g}Z_{c}}
\end{equation}
 $\delta _{\zeta }$ and $\delta _{\rho
}$ are defined as
\begin{eqnarray}
\delta _{\zeta }(g^{2}) &=&\left( \varepsilon-2\widehat{\gamma}
(g^{2})-\beta (g^{2})\frac{\partial }{\partial
g^{2}}-\alpha\gamma_{\alpha}(g^{2})\frac{\partial}{\partial\alpha}\right)
\delta \zeta
\label{d4bis} \\
\delta _{\rho }(g^{2}) &=&\left( \varepsilon-2\widehat{\gamma}
(g^{2})-\beta (g^{2})\frac{\partial }{\partial
g^{2}}-\alpha\gamma_{\alpha}(g^{2})\frac{\partial}{\partial\alpha}\right)
\delta \rho \label{d4}
\end{eqnarray}
where $\beta (g^{2})=\mu\frac{dg^{2}}{d\mu}$ is the usual running
of the coupling constant, in $d$ dimensions given by
\begin{equation}
\beta (g^{2})=-\varepsilon
g^{2}-\frac{22}{3}g^{2}\frac{g^{2}N}{16\pi
^{2}}-\frac{68}{3}g^{2}\left(\frac{g^{2}N}{16\pi
^{2}}\right)^{2}+\cdots \label{d4tris}
\end{equation}
while
$\gamma_{\alpha}(g^{2})=\frac{\mu}{\alpha}\frac{d\alpha}{d\mu}$
denotes the running of the gauge parameter $\alpha$. We do not
write the possible $\alpha$ dependence of the appearing
renormalization group functions; for the explicit calculations in
section 3.2, we will restrict ourselves to the Landau gauge.
Therefore, we also do not write down the explicit value of
$\gamma_{\alpha}(g^{2})$ since
$\alpha\gamma_{\alpha}(g^{2})\equiv0$ for $\alpha=0$.
\\\\ $\widehat{\gamma}(g^{2})$ denotes the anomalous dimension of the
sources $\omega$, $\tau$ and $L$. $\gamma(g^{2})$ and
$\widehat{\gamma}(g^{2})$ are related by
\begin{equation}\label{gammahat}
    \widehat{\gamma}(g^{2})=\frac{\varepsilon}{2}-\gamma(g^{2})
\end{equation}
and therefore, the equations (\ref{d4bis})-(\ref{d4}) can be
rewritten as
\begin{eqnarray}
\delta _{\zeta }(g^{2}) &=&\left( 2\gamma(g^{2})-\beta
(g^{2})\frac{\partial }{\partial
g^{2}}-\alpha\gamma_{\alpha}(g^{2})\frac{\partial}{\partial\alpha}\right)
\delta \zeta
\label{d4bisnew} \\
\delta _{\rho }(g^{2}) &=&\left( 2\gamma (g^{2})-\beta
(g^{2})\frac{\partial }{\partial
g^{2}}-\alpha\gamma_{\alpha}(g^{2})\frac{\partial}{\partial\alpha}\right)
\delta \rho \label{d4new}
\end{eqnarray}
Notice that in the equations (\ref{d3bis})-(\ref{d3}), the
parameter $\varepsilon$ has immediately been set equal to zero,
this is allowed because all considered quantities are finite for
$\varepsilon\rightarrow0$.
\\\\Since we have
introduced 2 novel parameters\footnote{In fact, only 1 novel
parameter is introduced, since $\rho=2\zeta$.}, we have a problem
of uniqueness. However, this can be solved by noticing that $\zeta
$ and $\rho $ can be chosen to be a function of $g^{2}$, such that
if $g^{2}$ runs according to (\ref{d4tris}), $\zeta(g^{2})$ and
$\rho(g^{2})$ will run according to (\ref{d3bis}), respectively
(\ref{d3}). Explicitly, $\zeta(g^{2})$ and $\rho(g^{2})$ are the
solution of the differential equations
\begin{eqnarray}
\left(\beta(g^{2}) \frac{d }{dg^{2}
}+\alpha\gamma_{\alpha}(g^{2})\frac{d}{d\alpha}\right)\zeta(g^{2})
&=&2\gamma (g^{2})\zeta(g^{2}) +\delta _{\zeta }(g^{2})
\label{t1a} \\
\left(\beta(g^{2}) \frac{d }{dg^{2}
}+\alpha\gamma_{\alpha}(g^{2})\frac{d}{d\alpha}\right)\rho(g^{2})
&=&2\gamma (g^{2})\rho(g^{2}) +\delta _{\rho }(g^{2}) \label{t1b}
\end{eqnarray}
The integration constants of the solution of
(\ref{t1a})-(\ref{t1b}) can be put to zero;this eliminates
independent parameters and assures multiplicative
renormalizability
\begin{eqnarray}
  \zeta(g^{2}) +\delta\zeta(g^{2},\varepsilon)&=& Z_{\zeta}(g^{2},\varepsilon)\zeta(g^{2}) \\
    \rho(g^{2}) +\delta\rho(g^{2},\varepsilon)&=& Z_{\rho}(g^{2},\varepsilon)\rho(g^{2})
\end{eqnarray}
Notice that the $n$-loop knowledge of $\zeta(g^{2})$ and
$\rho(g^{2})$ will need the $(n+1)$-loop knowledge of
$\beta(g^{2})$, $\gamma(g^{2})$, $\delta_{\zeta}(g^{2})$ and
$\delta_{\rho}(g^{2})$ \cite{Dudal:2003gu}. The generating
functional $\mathcal{W}(\omega,\tau,L)$, defined as
\begin{equation}
e^{i\mathcal{W}(\omega ,\tau ,L)}=\int [D\Phi ]e^{iS(\omega ,\tau
,L)} \label{d12}
\end{equation}
with $S(\omega,\tau,L)$ given by (\ref{d1}) and $\Phi $ denoting
the relevant fields, will now obey a homogeneous renormalization
group equation \cite{Verschelde:2001ia,Knecht:2001cc}. \\\\It is
not difficult to see that $\delta _{\zeta }(g^{2}),\delta
_{\rho}(g^{2})$ and $\zeta(g^{2}) ,\rho(g^{2}) $ will be of the
form
\begin{eqnarray}
\delta _{\zeta }(g^{2}) &=&\delta _{\zeta ,0}g^{2}+\delta _{\zeta
,1}g^{4}+\cdots
\label{d5} \\
\delta _{\rho }(g^{2}) &=&\delta _{\rho ,0}g^{2}+\delta _{\rho ,1}g^{4}+\cdots \\
\zeta(g^{2}) &=&\zeta _{0}+\zeta _{1}g^{2}+\cdots \\
\rho(g^{2}) &=&\rho _{0}+\rho_{1}g^{2}+\cdots
\end{eqnarray}
Taking the functional derivatives of $\mathcal{W}(\omega, \tau,
L)$ with respect to the sources $\omega^{a}$, $\tau^{a}$ and
$L^{a}$, we obtain a finite vacuum expectation value for the
composite operators, namely
\begin{eqnarray}
\left. \frac{\delta \mathcal{W}(\omega ,\tau ,L)}{\delta \omega
^{a}}\right|
_{\omega =0,\tau =0,L=0} &=&-g\left\langle f^{abc}\overline{c}%
^{b}c^{c}\right\rangle  \label{d13} \\
\left. \frac{\delta \mathcal{W}(\omega ,\tau ,L)}{\delta \tau
^{a}}\right|
_{\omega =0,\tau =0,L=0} &=&-\frac{g}{2}\left\langle f^{abc}\overline{c}^{b}%
\overline{c}^{c}\right\rangle \\
\left. \frac{\delta \mathcal{W}(\omega ,\tau ,L)}{\delta
L^{a}}\right|
_{\omega =0,\tau =0,L=0} &=&\frac{g}{2}\left\langle f^{abc}c^{b}c^{c}\right%
\rangle
\end{eqnarray}
Since the source terms appear quadratically, we seem to have lost
an energy interpretation. However, this can be dealt with by
introducing a pair of Hubbard-Stratonovich fields \mbox{($\sigma
^{a}$, $\overline{\sigma }^{a}$)} for
the $\tau L$ term, and a Hubbard-Stratonovich field $\phi ^{a}$ for the $%
\omega ^{2}$ term. For the functional generator
$\mathcal{W}(\omega ,\tau ,L) $, we then get
\begin{equation}
e^{i\mathcal{W}(\omega ,\tau ,L)}=\int [d\Phi ]e^{iS(\sigma ,\overline{%
\sigma },\phi )+i\int d^{4}x\left( \frac{\phi ^{a}}{g}\omega
^{a}+\frac{\sigma ^{a}}{g}L^{a}+\frac{\overline{\sigma }}{g}
^{a}\tau ^{a}\right) }  \label{d14}
\end{equation}
where the action $S(\sigma ,\overline{\sigma },\phi )$ is given by
\begin{eqnarray}\label{d15}
S(\sigma ,\overline{\sigma },\phi ) &=&S_{YM}+S_{GF+FP}+\int d^{4}x\left( -%
\frac{\sigma ^{a}\overline{\sigma }^{a}}{g^{2}\zeta }-\frac{\phi ^{a}\phi ^{a}}{%
2g^{2}\rho }+\frac{\overline{\sigma }^{a}}{2g\zeta}gf^{abc}c^{b}c^{c}%
\right.  \nonumber   \\
&-&\left. \frac{\sigma ^{a}}{2g\zeta }gf^{abc}\overline{c}^{b}%
\overline{c}^{c}-\frac{\phi ^{a}}{g\rho }gf^{abc}\overline{c}%
^{b}c^{c}-\frac{1}{2\rho }g^{2}\left(
f^{abc}\overline{c}^{b}c^{c}\right)
^{2}+\frac{1}{4\zeta }g^{2}f^{abc}c^{b}c^{c}f^{ade}\overline{c}^{d}\overline{c%
}^{e}\right)\nonumber\\
\end{eqnarray}
Notice also that in expression (\ref{d14}), the sources $\omega$ ,
$\tau$, $L$ are now linearly coupled to the fields $\phi$,
$\sigma$, $\osigma$, allowing thus for the correct energy
interpretation of the corresponding effective action. Taking the
functional derivatives gives the relations
\begin{eqnarray}
\label{hub1a}\left\langle \phi ^{a}\right\rangle &=&-g^{2}
\left\langle f^{abc}\overline{c}^{b}c^{c}\right\rangle  \label{d16} \\
\label{hub1b}\left\langle \sigma ^{a}\right\rangle
&=&\frac{g^{2}}{2}
\left\langle f^{abc}c^{b}c^{c}\right\rangle \\
\label{hub1c}\left\langle \overline{\sigma }^{a}\right\rangle
&=&-\frac{g^{2}}{2}\left\langle
f^{abc}\overline{c}^{b}\overline{c}^{c}\right\rangle
\end{eqnarray}
where all vacuum expectation values are now calculated with the
action (\ref {d15}). \\\\Summarizing, we have constructed a new,
multiplicatively renormalizable Yang-Mills action (\ref{d15}),
incorporating the possible existence of ghost condensates. As
such, if a non-trivial vacuum is favoured, we can perturb around a
more stable vacuum than the trivial one. The action (\ref{d15}) is
explicitly $NO$ invariant\footnote{The $NO$ variations of the
$\sigma^{a}$, $\osigma^{a}$ and $\phi^{a}$ fields can be
determined immediately from (\ref{hub1a})-(\ref{hub1c}).}. The
corresponding effective action $V(\sigma,\osigma,\phi)$ obeys a
homogeneous renormalization group equation. \\\\To find out
whether the groundstate effectively favours non-vanishing ghost
condensates, we will calculate the 1-loop effective potential. For
the sake of simplicity, we will restrict ourselves to the case of
$SU(2)$ Yang-Mills theories in the Landau gauge ($\alpha=0$). In
this context, we remark that one can prove that the vacuum energy
will be gauge parameter independent order by order. This proof is
completely analoguous to the one presented in \cite{Dudal:2003gu},
and is based on the fact that the derivative with respect to
$\alpha$ of the action (\ref{slco}) is a BRST exact form plus
terms proportional to the sources, which equal zero in the minima
of the effective potential. As such, the usual proof of gauge
parameter independence can be used \cite{book}.

\subsection{Calculation of the 1-loop effective potential for
$N=2$ in the Landau gauge} We will determine the effective
potential \cite{Coleman:jx} with the background field method
\cite{Jackiw:cv}. Let us define the $6\times6$ matrix
\begin{equation}
\mathcal{M}^{ab}=\left(
\begin{array}{cc}
-\frac{\sigma ^{c}\epsilon^{cab}}{\zeta} & \partial ^{2}\delta ^{ab}-\frac{%
\epsilon^{abc}\phi ^{c}}{\rho} \\
-\partial ^{2}\delta ^{ab}-\frac{\epsilon^{abc}\phi ^{c}}{\rho } & \frac{%
\overline{\sigma }^{c}\epsilon^{cab}}{\zeta}
\end{array}
\right)  \label{d17}
\end{equation}
where $\epsilon^{abc}$ are the structure constants of $SU(2)$.
Then the effective potential up to one loop is easily worked out,
yielding\footnote{We do not write the counterterms explicitly.}
\begin{equation}
V_{1}(\sigma ,\overline{\sigma },\phi )=\frac{\sigma ^{a}\overline{\sigma }%
^{a}}{g^{2}\zeta}+\frac{\phi ^{a}\phi ^{a}}{2g^{2}\rho}+\frac{i}{2}\ln \det \mathcal{M%
}^{ab}  \label{d18}
\end{equation}
or
\begin{equation}
V_{1}(\sigma ,\overline{\sigma },\phi )=\frac{\sigma ^{a}\overline{\sigma }%
^{a}}{g^{2}\zeta}+\frac{\phi ^{a}\phi ^{a}}{2g^{2}\rho}-\int
\frac{d^{d}k}{(2\pi
)^{d}}\ln \left( k^{6}+k^{2}\left(\frac{\sigma ^{a}\overline{\sigma }^{a}}{\zeta^{2} }+%
\frac{\phi ^{a}\phi ^{a}}{\rho^{2}
}\right)+\frac{\epsilon^{abc}\phi^{a}\sigma^{b}\osigma^{c}}{\rho\zeta^{2}}\right)
\label{d19}
\end{equation}
with $k$ Euclidean. \\\\We notice that the mass dimension 6
operator $\epsilon^{abc}\phi^{a}\sigma^{b}\osigma^{c}$ enters the
expression for the effective potential. We shall however show that
this operator plays no role in the determination of the minimum,
which is a solution of
\begin{eqnarray}\label{potjuist1}
    \left\{%
\begin{array}{lll}
    \frac{\partial V}{\partial\osigma^{a}} =\frac{\sigma^{a}}{g^{2}\zeta}-\int \frac{d^{d}k}{\left(2\pi\right)^{d}}
    \frac{k^{2}\frac{\sigma^{a}}{\zeta^{2}}+\frac{\epsilon^{abc}\phi^{b}\sigma^{c}}{\rho\zeta^{2}}}{k^{6}+k^{2}\left(\frac{\sigma ^{a}\overline{\sigma }^{a}}{\zeta^{2} }+%
\frac{\phi ^{a}\phi ^{a}}{\rho^{2}
}\right)+\frac{\epsilon^{abc}\phi^{a}\sigma^{b}\osigma^{c}}{\rho\zeta^{2}}} =0\\

    \frac{\partial V}{\partial\sigma^{a}} = \frac{\osigma^{a}}{g^{2}\zeta}-\int \frac{d^{d}k}{\left(2\pi\right)^{d}}
    \frac{k^{2}\frac{\osigma^{a}}{\zeta^{2}}+\frac{\epsilon^{abc}\phi^{c}\osigma^{b}}{\rho\zeta^{2}}}{k^{6}+k^{2}\left(\frac{\sigma ^{a}\overline{\sigma }^{a}}{\zeta^{2} }+%
\frac{\phi ^{a}\phi ^{a}}{\rho^{2}
}\right)+\frac{\epsilon^{abc}\phi^{a}\sigma^{b}\osigma^{c}}{\rho\zeta^{2}}}=0 \\

    \frac{\partial V}{\partial\phi^{a}} = \frac{\phi^{a}}{g^{2}\rho}-\int \frac{d^{d}k}{\left(2\pi\right)^{d}}
    \frac{2k^{2}\frac{\phi^{a}}{\rho^{2}}+\frac{\epsilon^{abc}\sigma^{b}\osigma^{c}}{\rho\zeta^{2}}}{k^{6}+k^{2}\left(\frac{\sigma ^{a}\overline{\sigma }^{a}}{\zeta^{2} }+%
\frac{\phi ^{a}\phi ^{a}}{\rho^{2}
}\right)+\frac{\epsilon^{abc}\phi^{a}\sigma^{b}\osigma^{c}}{\rho\zeta^{2}}}=0\\
\end{array}%
\right.
\end{eqnarray}
Let us assume that
$\left(\phi^{a}_{*},\sigma^{a}_{*},\osigma^{a}_{*}\right)$ is a
solution of (\ref{potjuist1}). Obviously, $\phi^{a}_{*}=0$,
$\sigma^{a}_{*}=0$, $\osigma^{a}_{*}=0$ is a solution,
corresponding with the trivial vacuum energy $E=0$.\\\\
Let us now assume that at least one of the field configurations is
non-zero. If it occurs that
$\sigma^{a}_{*}=\osigma^{a}_{*}=(0,0,0)$, then necessarily
$\phi^{a}_{*}\neq(0,0,0)$ and it can be immediately checked that
the equations (\ref{potjuist1}) are reduced to
\begin{equation}\label{potjuist2}
    \frac{1}{g^{2}\zeta}-\frac{1}{{\zeta^{2}}}\int
\frac{d^{d}k}{\left(2\pi\right)^{d}}
    \frac{1}{k^{4}+\left(\frac{\sigma ^{a}_{*}\overline{\sigma }^{a}_{*}}{\zeta^{2} }+%
\frac{\phi ^{a}_{*}\phi ^{a}_{*}}{\rho^{2} }\right)} =0
\end{equation}
Next, we consider the case that $\sigma^{a}_{*}\neq(0,0,0)$ and/or
$\osigma^{a}_{*}\neq(0,0,0)$. Without loss of generality, we can
consider $\sigma^{a}_{*}\neq(0,0,0)$. Consider then the first
equation of (\ref{potjuist1}).
\begin{equation}\label{potjuist3}
    \frac{\sigma^{a}_{*}}{g^{2}\zeta}-\int \frac{d^{d}k}{\left(2\pi\right)^{d}}
    \frac{k^{2}\frac{\sigma^{a}_{*}}{\zeta^{2}}+\frac{\epsilon^{abc}\phi^{b}_{*}\sigma^{c}_{*}}{\rho\zeta^{2}}}{k^{6}+k^{2}\left(\frac{\sigma ^{a}_{*}\overline{\sigma }^{a}_{*}}{\zeta^{2} }+%
\frac{\phi ^{a}_{*}\phi ^{a}_{*}}{\rho^{2}
}\right)+\frac{\epsilon^{abc}\phi^{a}_{*}\sigma^{b}_{*}\osigma^{c}_{*}}{\rho\zeta^{2}}}
=0
\end{equation}
By contracting the above equation with $\sigma^{a}_{*}$, we find
\begin{equation}\label{potjuist4}
\frac{\sigma^{a}_{*}\sigma^{a}_{*}}{g^{2}\zeta}-\int
\frac{d^{d}k}{\left(2\pi\right)^{d}}
    \frac{k^{2}\frac{\sigma^{a}_{*}\sigma^{a}_{*}}{\zeta^{2}}}{k^{6}+k^{2}\left(\frac{\sigma ^{a}_{*}\overline{\sigma }^{a}_{*}}{\zeta^{2} }+%
\frac{\phi ^{a}_{*}\phi ^{a}_{*}}{\rho^{2}
}\right)+\frac{\epsilon^{abc}\phi^{a}_{*}\sigma^{b}_{*}\osigma^{c}_{*}}{\rho\zeta^{2}}}
=0
\end{equation}
or, since $\sigma^{a}_{*}\sigma^{a}_{*}\neq0$
\begin{equation}\label{potjuist5}
\frac{1}{g^{2}\zeta}-\int \frac{d^{d}k}{\left(2\pi\right)^{d}}
    \frac{\frac{k^{2}}{{\zeta^{2}}}}{k^{6}+k^{2}\left(\frac{\sigma ^{a}_{*}\overline{\sigma }^{a}_{*}}{\zeta^{2} }+%
\frac{\phi ^{a}_{*}\phi ^{a}_{*}}{\rho^{2}
}\right)+\frac{\epsilon^{abc}\phi^{a}_{*}\sigma^{b}_{*}\osigma^{c}_{*}}{\rho\zeta^{2}}}
=0
\end{equation}
Inserting (\ref{potjuist5}) into (\ref{potjuist3}), one learns
that
\begin{equation}\label{potjuist6}
    \frac{\epsilon^{abc}\phi^{b}_{*}\sigma^{c}_{*}}{\rho\zeta^{2}}\int \frac{d^{d}k}{\left(2\pi\right)^{d}}
    \frac{1}{k^{6}+k^{2}\left(\frac{\sigma ^{a}_{*}\overline{\sigma }^{a}_{*}}{\zeta^{2} }+%
\frac{\phi ^{a}_{*}\phi ^{a}_{*}}{\rho^{2}
}\right)+\frac{\epsilon^{abc}\phi^{a}_{*}\sigma^{b}_{*}\osigma^{c}_{*}}{\rho\zeta^{2}}}
=0
\end{equation}
Notice that the integral in (\ref{potjuist6}) is UV finite. If the
integral of (\ref{potjuist6}) is non-vanishing, we must have that
\begin{equation}\label{potjuist7}
    \epsilon^{abc}\phi^{b}_{*}\sigma^{c}_{*}=0
\end{equation}
Evidently, we then also have that
\begin{equation}\label{potjuist8}
    \epsilon^{abc}\phi^{a}_{*}\sigma^{b}_{*}\osigma^{c}_{*}=0
\end{equation}
Expression (\ref{potjuist5}) can then also be combined with the
second and third equation of (\ref{potjuist1}) to show that
\begin{equation}\label{potjuist9}
    \epsilon^{abc}\phi^{b}_{*}\osigma^{c}_{*}=0
\end{equation}
and
\begin{equation}\label{potjuist10}
    \epsilon^{abc}\sigma^{b}_{*}\osigma^{c}_{*}=0
\end{equation}
Henceforth, we conclude that all contributions coming from the
dimension 6 operator $\epsilon^{abc}\phi^{b}\sigma^{c}\osigma^{a}$
are in fact not relevant for the determination of the minimum
configuration
$\left(\phi^{a}_{*},\sigma^{a}_{*},\osigma^{a}_{*}\right)$. It is
sufficient to solve the following gap equation to search for the
non-trivial minimum
\begin{equation}\label{potjuist10}
\frac{1}{g^{2}\zeta}-\frac{1}{{\zeta^{2}}}\int
\frac{d^{d}k}{\left(2\pi\right)^{d}}
    \frac{1}{k^{4}+\left(\frac{\sigma ^{a}\overline{\sigma }^{a}}{\zeta^{2} }+%
\frac{\phi ^{a}\phi ^{a}}{\rho^{2} }\right)} =0
\end{equation}
In fact, this is the gap equation corresponding to the
minimization of the potential (\ref{d19}) with
$\epsilon^{abc}\phi^{a}\sigma^{b}\osigma^{c}$ put equal to zero
from the beginning, in which case the 1-loop potential reduces to
\begin{eqnarray}\label{d19tris}
V_{1}(\sigma ,\overline{\sigma },\phi )^{\epsilon^{abc}\phi^{a}\sigma^{b}\osigma^{c}=0} &=&\frac{\sigma ^{a}\overline{%
\sigma
}^{a}}{g^{2}\zeta_{0}}\left(1-\frac{\zeta_{1}}{\zeta_{0}}g^{2}\right)+\frac{\phi
^{a}\phi
^{a}}{2\rho_{0}g^{2}}\left(1-\frac{\rho_{1}}{\rho_{0}}g^{2}\right)
\nonumber\\&+&\frac{1}{32\pi ^{2}}%
\left( \frac{\sigma ^{a}\overline{\sigma }^{a}}{\zeta
_{0}^{2}}+\frac{\phi
^{a}\phi ^{a}}{\rho _{0}^{2}}\right)\left( \ln \frac{\frac{\sigma ^{a}\overline{\sigma }^{a}}{\zeta _{0}^{2}%
}+\frac{\phi ^{a}\phi ^{a}}{\rho _{0}^{2}}}{\overline{\mu
}^{4}}-3\right)
\end{eqnarray}
Moreover, we have explicitly verified that that the potential
$V_{1}$, for $\epsilon^{abc}\phi^{a}\sigma^{b}\osigma^{c}\neq0$,
does in fact admit the solution
$\epsilon^{abc}\phi^{a}\sigma^{b}\osigma^{c}=0$ for the
minimum.\\\\
It remains to show that the integral of (\ref{potjuist6}) is
non-vanishing for a non-trivial vacuum configuration
($E_{\textrm{\tiny{vac}}}\neq0$). We define
\begin{eqnarray}
a&=&\frac{\sigma ^{a}_{*}\overline{\sigma }^{a}_{*}}{\zeta^{2} }+%
\frac{\phi ^{a}_{*}\phi ^{a}_{*}}{\rho^{2} }\nonumber\\
b&=&\frac{\epsilon^{abc}\phi^{a}_{*}\sigma^{b}_{*}\osigma^{c}_{*}}{\rho\zeta^{2}}
\end{eqnarray}
and consider the integral
\begin{equation}\label{potje1}
    \int
    \frac{d^{4}k}{\left(2\pi\right)^{4}}\frac{1}{k^{6}+ak^{2}+b}=\int
    \frac{d\Omega}{\left(2\pi\right)^{4}}\int_{0}^{\infty} \frac{k^{3}dk}{k^{6}+ak^{2}+b}
\end{equation}
For $a=0$ and $b=0$, (\ref{potje1}) is vanishing, but then we also
have that $E_{\textrm{\tiny{vac}}}=0$.  \\Via the substitution
$x=k^{2}$, one finds
\begin{equation}\label{potje2}
\int \frac{k^{3}dk}{k^{6}+ak^{2}+b}=\frac{1}{2}\int_{0}^{\infty}
\frac{xdx}{x^{3}+ax+b}
\end{equation}
This integral is always positive for $a>0$. For $b=0$, this is
immediately clear. For $b\neq0$, we perform a partial integration
to find
\begin{equation}\label{potje3}
\frac{1}{2}\int_{0}^{\infty}
\frac{xdx}{x^{3}+ax+b}=\left.\frac{x^{2}}{4\left(x^{3}+ax+b\right)}\right|_{0}^{\infty}
    +\frac{1}{4}\int_{0}^{\infty}\frac{\left(3x^{2}+a\right)x^{2}}{\left(x^{3}+ax+b\right)^{2}}dx
    =\frac{1}{4}\int_{0}^{\infty}\frac{\left(3x^{2}+a\right)x^{2}}{\left(x^{3}+ax+b\right)^{2}}dx
\end{equation}
For $a>0$, the integral (\ref{potje3}) is also positive. Consider
now the function $F(a,b)$, defined by
\begin{equation}\label{potje4}
    F(a,b)=\int_{0}^{\infty}\frac{xdx}{x^{3}+ax+b}
\end{equation}
We already know that, for $a>0$ and fixed $b=b_{*}$,
$F(a,b_{*})>0$. Furthermore
\begin{equation}\label{potje5}
    \frac{\partial F(a,b)}{\partial
    a}=-\int_{0}^{\infty}\frac{x^{2}dx}{\left(x3+ax+b\right)^{2}}<0
\end{equation}
meaning that the function $F(a,b_{*})$ decreases for increasing
$a$. Assuming that $F(a,b)$ has a zero at $(a_{0},b_{0})$, then we
should have that $F(a,b_{0})$ becomes more negative as $a$
increases, which contradicts the fact that $F(a,b_{0})>0$ for
$a>0$. Therefore, the function $F(a,b)$ cannot become zero and the
integral in (\ref{potjuist6}) never vanishes for a non-trivial
vacuum configuration.
\\\\It remains to calculate $\zeta_{0}$, $\zeta_{1}$, $\rho_{0}$ and
$\rho_{1}$. One finds (see the Appendix B)
\begin{eqnarray}
\label{t3a}\delta\zeta&=&-\frac{g^{2}}{8\pi^{2}}\frac{1}{\varepsilon}
+\frac{g^{4}}{(16\pi^{2})^{2}}\left(\frac{1}{2\varepsilon}+\frac{6}{\varepsilon^{2}}\right)+\cdots\\
\label{t3b}\delta\rho&=&-\frac{g^{2}}{4\pi^{2}}\frac{1}{\varepsilon}
+\frac{g^{4}}{(16\pi^{2})^{2}}\left(\frac{1}{\varepsilon}+\frac{12}{\varepsilon^{2}}\right)+\cdots
\end{eqnarray}
Since in the Landau gauge $Z_{2}=1$ and
$Z_{c}=Z_{g}^{-1}Z_{A}^{-1/2}$ (see e.g. \cite{Dudal:2002pq}), we
have
\begin{equation}\label{runningL}
    \gamma(g^{2})=\frac{1}{2}\mu\frac{d}{d\mu}\ln Z_{A}\equiv\gamma_{A}(g^{2})
\end{equation}
where $\gamma_{A}(g^{2})$ is the anomalous dimension of the gluon
field, given by \cite{Larin:tp,Gracey:2002yt}
\begin{equation}\label{runningL2}
    \gamma_{A}(g^{2})=-\frac{13}{6}\frac{g^{2}N}{16\pi^{2}}-\frac{59}{8}\left(\frac{g^{2}N}{16\pi^{2}}\right)^{2}+\ldots
\end{equation}
Henceforth, we find for (\ref{d4bisnew})-(\ref{d4new})
\begin{eqnarray}
\delta _{\zeta }(g^{2})
&=&-\frac{g^{2}}{8\pi^{2}}+\frac{g^{4}}{256\pi^{4}}+\cdots
\label{delta1} \\
\delta _{\rho }(g^{2})
&=&-\frac{g^{2}}{4\pi^{2}}+\frac{g^{4}}{128\pi^{4}}+\cdots
\label{delta1bis}
\end{eqnarray}
Another good internal check of the calculations\footnote{See also
the Appendix B.} is that the renormalization group functions
(\ref{delta1})-(\ref{delta1bis}) are indeed finite. \\\\Finally,
solving the equations (\ref{t1a})-(\ref{t1b}) leads to
\begin{eqnarray}
\label{t4a}\zeta_{0}&=&-\frac{3}{13}\\
\label{t4b}\rho_{0}&=&-\frac{6}{13}\\
\label{t4c}\zeta_{1}&=&-\frac{95}{624\pi^{2}}\\
\label{t4c}\rho_{1}&=&-\frac{95}{312\pi^{2}}
\end{eqnarray}
We indeed find that $\rho=2\zeta$. We already knew this from the
$NO$ invariance (see the \mbox{Appendix A}), and we find that the
$\overline{MS}$ scheme preserves this symmetry. It can also be
understood from a diagrammatical point of view. Consider
(\ref{d1}), first with only the source $\omega$ connected, and
subsequently with only the sources $\tau$, $L$ connected. For each
diagram giving a divergence proportional to $\omega^{2}$ in the
former case, there exists a similar diagram giving a divergence
proportional to $\tau L$ in the latter case. More precisely, when
the appropriate symmetry factor is taken into account, it will
hold that
\begin{equation}
\delta \rho =2\delta \zeta  \label{d11}
\end{equation}
Combining this with (\ref{d4bis})-(\ref{d4}) and
(\ref{t1a})-(\ref{t1b}), precisely gives the relation (\ref{rel}).
\\\\Notice that, due to the identity (\ref{rel}), the effective potential
$V(\sigma,\osigma,\phi)$ of (\ref{d18}) can be written in terms of
2 combinations of the fields $\sigma$, $\osigma$ and $\phi$,
namely
\begin{eqnarray}\label{t10}
    \chi^{2}&=&\sigma^{a}\osigma^{a}+\frac{\phi^{a}\phi^{a}}{4}\nonumber\\
    \widehat{\chi}&=&\epsilon^{abc}\phi^{a}\sigma^{b}\osigma^{c}
\end{eqnarray}
As we have shown, $\widehat{\chi}$ does not influence the value of
the minimum. So, it is sufficient to consider the potential with
$\widehat{\chi}=0$. (\ref{d19tris}) then becomes
\begin{equation}\label{RGE3}
    V_{1}(\chi)^{\widehat{\chi}=0}=\frac{\chi^{2}}{g^{2}\zeta_{0}}\left(1-\frac{\zeta_{1}}{\zeta_{0}}g^{2}\right)
    +\frac{1}{32\pi^{2}}\frac{\chi^{2}}{\zeta_{0}^{2}}\left(\ln\frac{\chi^{2}}{\zeta_{0}^{2}\overline{\mu}^{4}}-3\right)
\end{equation}
Recalling (\ref{d}) and (\ref{d16}), we find
\begin{eqnarray}
\delta \phi &=&-2\sigma   \label{d22trisa}\\
\delta \sigma &=&0 \label{d22trisb}\\
\delta \overline{\sigma } &=&\phi\label{d22trisc}
\end{eqnarray}
Consequently
\begin{eqnarray}
\delta
\chi^{2}&=&\phi^{a}\sigma^{a}+\frac{(2\phi^{a})(-2\sigma^{a})}{4}=0\nonumber\\
\delta\widehat{\chi}&=&0 \label{d23}
\end{eqnarray}
A similar conclusion exists for $\odelta$ and $\delta_{FP}$. Said
otherwise, $\chi$ and $\widehat{\chi}$ are $SL(2,R)$ invariants.
Let us make a comparison with the effective potential
$V(\varphi^{2})$ of the $O(N)$ vector model with field
$\varphi=(\varphi_{1},\ldots,\varphi_{N})$. This potential is a
function of the $O(N)$ invariant norm
$\varphi^{2}=\varphi_{1}^{2}+\cdots+\varphi_{N}^{2}$. Choosing a
certain direction for $\varphi$ breaks the $O(N)$ invariance. In
the present case, choosing a certain direction for $\chi$ breaks
the $SL(2,R)$ symmetry. However, the situation with the ghost
condensates is a bit more complicated than a simple breakdown of
the $SL(2,R)$. \\\\Before we come to the discussion of the
symmetry breaking, let us calculate the minima of (\ref{RGE3}). We
can use the renormalization group equation to sum leading
logarithms and put $\overline{\mu}^{4}=\frac{\chi^{2}}{\zeta
_{0}^{2}}$. The equation of motion, $\frac{dV}{d\chi}=0$, has,
next to the perturbative one $\chi=0$, which corresponds to a
local maximum, a non-trivial solution, given by
\begin{eqnarray}
  \label{opl1}\left.\frac{g^{2}N}{16\pi^{2}}\right|_{N=2} &=& \frac{9}{28}\approx0.321
\end{eqnarray}
where it is understood that $g^{2}\equiv
g^{2}(\overline{\mu}=\sqrt{\chi}/|\zeta_{0}|)$. Using the 1-loop
expression
\begin{equation}\label{opl4}
    g^{2}(\overline{\mu})=\frac{3}{11N}\frac{1}{\ln\frac{\overline{\mu}^{2}}{\Lambda_{\overline{MS}}^{2}}}
\end{equation}
we obtain
\begin{eqnarray}
  \label{opl2}\chi_{\tiny\textrm{vac}} &=& 0.539\Lambda_{\overline{MS}}^{2} \\
  \label{opl3}E_{\tiny\textrm{vac}} &=& -0.017\Lambda_{\overline{MS}}^{4}
\end{eqnarray}
>From (\ref{opl1}), it follows that the expansion parameter is
relatively small. A qualitatively meaningful minimum,
(\ref{opl2}), is thus retrieved. The resulting vacuum energy
(\ref{opl3}) is negative, implying that the ground state favours
the formation of the ghost condensates.

\sect{Non-trivial vacuum configurations and dynamical breaking of
the $NO$ symmetry} In this section, we discuss the consequences
for the $NO$ symmetry of a non-trivial vacuum expectation value of
the ghost operators $f^{abc}c^{b}c^{c}$, $f^{abc}\oc^{b}\oc^{c}$
and/or $f^{abc}\oc^{b}c^{c}$. The arguments are general and
applicable for all $N$ and for all choices of the Curci-Ferrari
gauge parameter $\alpha$.

\subsection{BCS, Overhauser or a combination of both?}
Since the action (\ref{d15}) is $NO$ invariant, each possible
vacuum state can be transformed into another under the action of
the $NO$ symmetry. A special choice of a possible vacuum is the
pure Overhauser vacuum, determined by\footnote{Without loss of
generality, we can put $\phi^{a}$ in the 3-direction.}
\begin{eqnarray}\label{OV}
    \left\{%
\begin{array}{ll}
    \phi^{a} = \phi_{\tiny{\textrm{vac}}}\delta^{a3} \textrm{ with } \phi_{\tiny{\textrm{vac}}}=2\chi_{\tiny{\textrm{vac}}} \\
    \sigma^{a} = \osigma^{a}=0 \\
\end{array}%
\right.
\end{eqnarray}
Then two of the $SL(2,R)$ generators ($\delta$ and $\odelta$) are
dynamically broken since
\begin{eqnarray}
  \left\langle\delta\osigma\right\rangle=-\left\langle\odelta\sigma\right\rangle=
  \left\langle\phi\right\rangle\neq0
\end{eqnarray}
The ghost number symmetry $\delta_{FP}$ is unbroken, just as the
BRST symmetry $s$, since no operator $\mathcal{F}$ exists with
$\left\langle
s\mathcal{F}\right\rangle=\left\langle\phi\right\rangle$. In fact,
setting
\begin{eqnarray}\label{brst1}
    \phi^{a}&=&\phi_{\tiny{\textrm{vac}}}\delta^{a3}+\widetilde{\phi}^{a}\textrm{ with }\left\langle\widetilde{\phi}^{a}\right\rangle=0 \\
    s\widetilde{\phi}^{a}&=&-g^{2}s\left(f^{abc}\oc^{b}c^{c}\right)
\end{eqnarray}
it is immediately verified that the action
\begin{eqnarray}\label{brst2}
S(\sigma ,\overline{\sigma },\widetilde{\phi} ) &=&S_{YM}+S_{GF+FP}+\int d^{4}x\left( -%
\frac{\sigma ^{a}\overline{\sigma }^{a}}{g^{2}\zeta }-\frac{\phi_{\tiny{\textrm{vac}}}^{2}}{%
2g^{2}\rho }-\frac{\widetilde{\phi} ^{3}\phi_{\tiny{\textrm{vac}}}}{%
g^{2}\rho }-\frac{\widetilde{\phi} ^{a}\widetilde{\phi} ^{a}}{%
2g^{2}\rho }+\frac{\overline{\sigma }^{a}}{2g\zeta}gf^{abc}c^{b}c^{c}%
\right.  \nonumber   \\
&-&\left. \frac{\sigma ^{a}}{2g\zeta }gf^{abc}\overline{c}^{b}%
\overline{c}^{c}-\frac{\widetilde{\phi} ^{a}}{g\rho }gf^{abc}\overline{c}%
^{b}c^{c}-\frac{\phi_{\tiny{\textrm{vac}}}}{g\rho }gf^{3bc}\overline{c}%
^{b}c^{c}\right.\nonumber\\&-&\left.\frac{1}{2\rho }g^{2}\left(
f^{abc}\overline{c}^{b}c^{c}\right)
^{2}+\frac{1}{4\zeta }g^{2}f^{abc}c^{b}c^{c}f^{ade}\overline{c}^{d}\overline{c%
}^{e}\right)\nonumber\\
\end{eqnarray}
obeys
\begin{equation}\label{brst3}
    sS(\sigma ,\overline{\sigma },\widetilde{\phi} )=0
\end{equation}
while evidently
\begin{equation}\label{brst4}
    s^{2}=0
\end{equation}
We focus on the ghost number and BRST symmetry because these are
the key ingredients for the definition of a physical subspace, to
have a quartet mechanism, etc.; see e.g. \cite{Kugo:gm}.
\\\\
For vacua other than the pure Overhauser case, problems can arise
concerning the BRST and/or the ghost number symmetry. Consider for
example the pure BCS vacuum
\begin{eqnarray}\label{pureBCS}
    \left\{%
\begin{array}{lll}
    \phi^{a} = 0 \\
    \sigma^{a} = b\chi_{\tiny{\textrm{vac}}}\delta^{a3}\\
    \osigma^{a}=\overline{b}\chi_{\tiny{\textrm{vac}}}\delta^{a3} \\
    \end{array}%
\right.
\end{eqnarray}
where $b$ and $\overline{b}$ are a pair of Faddeev-Popov
conjugated constants ($b\overline{b}=1$). In this vacuum,
$\left\langle f^{abc}c^{b}c^{c}\right\rangle\neq0$, while
$sc^{a}=\frac{g}{2}f^{abc}c^{b}c^{c}$, so we can expect a problem
with the BRST transformation. Things can even be made worse, since
also vacua where $\sigma^{a}$ and $\osigma^{a}$ get a different
value (up to the ghost number, which is $2$, respectively $-2$),
are allowed. In
this case, the ghost number symmetry $\delta_{FP}$ is also broken.\\\\
It seems that the existence of the ghost condensates, different
from the Overhauser channel, could cause serious problems. A
pragmatic solution would be to simply choose the Overhauser
vacuum, since one always has to choose a specific vacuum to work
with. However, this is not very satisfactory. The other vacua are
in principle as 'good' as the Overhauser one. \\\\Let us try to
formulate a solution to the problem of the possible BRST/ghost
number symmetry breakdown. Let $\left|\Omega\right\rangle$ be the
Overhauser vacuum, and $\left|\widetilde{\Omega}\right\rangle$ any
other vacuum. As already said, a certain $NO$ transformation
$\mathcal{U}$ exists, so that
\begin{equation}\label{vac1}
       \left|\widetilde{\Omega}\right\rangle=\mathcal{U}\left|\Omega\right\rangle
\end{equation}
Let $Q_{BRST}$, $\overline{Q}_{BRST}$, $Q_{FP}$, $Q_{\delta}$ and
$Q_{\overline{\delta}}$ be the charges corresponding to
respectively $s$, $\os$, $\delta_{FP}$, $\delta$ and
$\overline{\delta}$. We know that
\begin{eqnarray}
      \label{vac2a}Q_{BRST}\left|\Omega\right\rangle &=& 0 \\
      \label{vac2c}Q_{FP}\left|\Omega\right\rangle&=& 0
\end{eqnarray}
With the relations (\ref{vac1})-(\ref{vac2c}), it is possible to
define new charges\footnote{As it is well known, the generators of
a symmetry form an adjoint representation.}
\begin{eqnarray}
      \label{vac3a}\widetilde{Q}_{BRST} &=& \mathcal{U}Q_{BRST}\mathcal{U}^{-1} \\
      \label{vac3b}\widetilde{\overline{Q}}_{BRST} &=& \mathcal{U}\overline{Q}_{BRST}\mathcal{U}^{-1} \\
      \label{vac3c}\widetilde{Q}_{FP}&=&\mathcal{U}Q_{FP}\mathcal{U}^{-1}\\
      \label{vac3d}\widetilde{Q}_{\delta}&=&\mathcal{U}Q_{\delta}\mathcal{U}^{-1}\\
     \label{vac3e}\widetilde{Q}_{\overline{\delta}}&=&\mathcal{U}Q_{\overline{\delta}}\mathcal{U}^{-1}
\end{eqnarray}
Since this is merely a redefinition of its generators, the new
charges (\ref{vac3a})-(\ref{vac3e}) are evidently still obeying
the $NO$ algebra (\ref{no}). By construction, we
have\footnote{$\widetilde{Q}_{\delta}$ for example will be a
broken generator. If not, one has
$Q_{\delta}\left|\Omega\right\rangle=0$, a contradiction.}
\begin{eqnarray}
      \label{vac4a}\widetilde{Q}_{BRST}\left|\widetilde{\Omega}\right\rangle &=& 0 \\
      \label{vac4c}\widetilde{Q}_{FP}\left|\widetilde{\Omega}\right\rangle&=& 0
\end{eqnarray}
As such, we have in \emph{any} vacuum $\widetilde{\Omega}$ the
concept of a nilpotent operator $\widetilde{Q}_{BRST}$.
Furthermore, the physical states
$\left|\widetilde{\textrm{phys}}\right\rangle$ are those wherefore
\begin{eqnarray}\label{vac5a}
    \label{vac5a}\widetilde{Q}_{BRST}\left|\widetilde{\textrm{phys}}\right\rangle&=&0\\
    \label{vac5b}\left|\widetilde{\textrm{phys}}\right\rangle&\neq&\widetilde{Q}_{BRST}\left|\ldots\right\rangle\\
    \label{vac5c}\widetilde{Q}_{FP}\left|\widetilde{\textrm{phys}}\right\rangle&=&0
\end{eqnarray}
and are connected to the physical states of the Overhauser case
through
\begin{equation}\label{physa}
    \left|\widetilde{\textrm{phys}}\right\rangle=\mathcal{U}\left|\textrm{phys}\right\rangle
\end{equation}
The conclusion is that in any vacuum, the concept of a
Faddeev-Popov symmetry exists, just as a nilpotent BRST
transformation. The mere difference is that the functional form of
these operators is no longer the usual one (\ref{s}). But in
principle, the $\sim$ generators are as good as the original ones
to perform the Kugo-Ojima formalism, since this is based on
algebraic properties \cite{Kugo:gm}. The $NO$ can thus be used to
define the physical subspace $\mathcal{H}_{\tiny{\textrm{phys}}}$
of the total Hilbert space $\mathcal{H}$ of all possible states.
The action of the $NO$ rotates $\mathcal{H}$, whereby '$Q_{BRST}$
physical' states $\left|\textrm{phys}\right\rangle$ are rotated
into '$\widetilde{Q}_{BRST}$ physical' states
$\left|\widetilde{\textrm{phys}}\right\rangle\equiv\mathcal{U}\left|\textrm{phys}\right\rangle$.
\\\\ Since we have to choose a certain vacuum, we assume for the
rest of the article that we are in the Overhauser vacuum, the most
obvious choice. Notice that this does not imply that we can simply
put the sources $L^{a}$ and $\tau^{a}$ equal to zero from the
beginning. This corresponds to the ghost condensation studied in
the context of the Maximal Abelian Gauge, originated in
\cite{Schaden:1999ew,Schaden:2000fv,Schaden:2001xu,Kondo:2000ey}.
Analogously, setting $\omega^{a}$ equal to zero from the
beginning, corresponds to the BCS channel as originally studied in
\cite{Lemes:2002ey,Lemes:2002jv,Lemes:2002rc}.

\subsection{Global color symmetry}
A non-vanishing vacuum expectation value for the color charged
field $\phi^{a}$ seems to spoil the global color symmetry, i.e.
the global $SU(N)$ invariance. However, it can be argued that this
global color symmetry breaking is located in the unphysical sector
of the Hilbert space. According to \cite{book2,Kugo:gm}, the
conserved, global $SU(N)$ current is given by
\begin{equation}\label{curr}
    \mathcal{J}_{\mu}^{a}=\partial_{\nu}F_{\mu\nu}^{a}+\{Q_{BRST},D_{\mu}^{ab}\oc^{b}\}
\end{equation}
while the corresponding color charge reads
\begin{equation}\label{charge}
    \mathcal{Q}^{a}=\int d^{3}x\partial_{i}F_{0i}^{a}+\int d^{3}x\{Q_{BRST},D_{0}^{ab}\oc^{b}\}
\end{equation}
The current (\ref{curr}) is the same in comparison with the one
given by the usual Yang-Mills Lagrangian (i.e. without any
condensate); this is immediately verified since the action
(\ref{d15}) does not contain any new terms with derivatives of the
fields.\\\\The first term of (\ref{charge}) is either ill-defined
due to massless particles in its spectrum, or zero as a volume
integral of a total divergence \cite{Alkofer:2000wg}. Thus, if no
massless particles show up (i.e. gluons are massive),
(\ref{charge}) reduces to a BRST exact form
\begin{equation}\label{charge2}
    \mathcal{Q}^{a}=\int d^{3}x\{Q_{BRST},D_{0}^{ab}\oc^{b}\}
\end{equation}
Henceforth, this color breaking should not be observed in the
physical subspace of the Hilbert space, see e.g.
\cite{Alkofer:2000wg} and references therein.
\\\\The required absence of massless particles is assured if the
gluons are no longer massless. This is realized by another
condensate of mass dimension 2, namely $\frac{1}{2}\left\langle
A^{2}\right\rangle$ in the case of the Landau gauge. This
condensate also lowers the vacuum energy and gives rise to a
dynamical gluon mass, as was shown in
\cite{Verschelde:2001ia,Dudal:2003vv}. Also lattice simulations
support a dynamical gluon mass \cite{Langfeld:2001cz,
Alexandrou:2001fh}. The generalization to the Curci-Ferrari gauge
was discussed in \cite{Dudal:2003gu}. \\\\A rather subtle point in
the foregoing is that the well-definedness of (\ref{charge2})
should be assured.

\subsection{Absence of Goldstone excitations}
The conserved current corresponding to the $\delta$ invariance is
given by
\begin{eqnarray}
  \label{k1}k_{\mu} = c^{a}D_{\mu}^{ab}c^{b}+\frac{1}{2}gf^{abc}A_{\mu}^{a}c^{b}c^{c}=s\left(c^{a}A_{\mu}^{a}\right)
\end{eqnarray}
An analogous expression can be derived for the $\overline{\delta}$
current
\begin{eqnarray}
  \label{k2}\overline{k}_{\mu}
  =\os\left(\overline{c}^{a}A_{\mu}^{a}\right)
\end{eqnarray}
If these continuous $\delta$ and $\overline{\delta}$ symmetries
are broken, massless Goldstone states should appear, according to
the Goldstone theorem. However, since the currents are (anti-)BRST
exact, those Goldstone bosons will be part of a BRST quartet, and
as such decouple from the physical spectrum due to the quartet
mechanism \cite{Kugo:gm}. The argument is analogous to the one
given in \cite{Schaden:1999ew,Schaden:2000fv,Schaden:2001xu} to
explain why there are no physical Goldstone particles present in
the case of $SU(2)$ Yang-Mills in the Maximal Abelian gauge, due
to the appearance of the condensate
$\left\langle\epsilon^{3ab}\overline{c}^{a}c^{b}\right\rangle$.

\sect{Inclusion of matter fields} So far, we have considered pure
Yang-Mills theories, i.e. without matter fields. The present
analysis can be nevertheless straightforwardly extended to the
case with quarks included. This is accomplished by adding to the
pure Yang-Mills action $S_{YM}$ the quark contribution $S_{m}$,
given by
\begin{equation}\label{q1}
    S_{m}=\int d^{4}x
    \overline{\psi}^{iI}i\gamma^{\mu}D_{\mu}^{IJ}\psi^{iJ}
\end{equation}
with
\begin{equation}\label{q2}
    D_{\mu}^{IJ}=\partial_{\mu}\delta^{IJ}-igA_{\mu}^{a}T^{aIJ}
\end{equation}
The $T^{aIJ}$ are the generators of the fundamental representation
of $SU(N)$, while $D_{\mu}^{IJ}$ is the corresponding covariant
derivative. The index $i$ labels the number of flavours $(1\leq
i\leq N_{f})$.\\\\The action of the $NO$ transformation on the
fermion fields is defined as follows
\begin{eqnarray}
  s\psi^{iI} &=& -igc^{a}T^{aIJ}\psi^{iJ} \\
  s\overline{\psi}^{iI} &=& -ig\overline{\psi}^{iJ}T^{aJI}c^{a}  \\
  \nonumber\\
 \os\psi^{iI} &=& -ig\oc^{a}T^{aIJ}\psi^{iJ} \\
  \os\overline{\psi}^{iI} &=& -ig\overline{\psi}^{iJ}T^{aJI}\oc^{a}  \\
\nonumber\\
    \delta \psi&=&\odelta\psi=\delta_{FP}\psi=0\\
    \nonumber\\
    \delta \overline{\psi}&=&\odelta\overline{\psi}=\delta_{FP}\overline{\psi}=0
\end{eqnarray}
Then it is easily checked that the algebra structure (\ref{no}) is
maintained, while the full action
\begin{equation}\label{q3}
    S=S_{YM}+S_{m}+S_{GF+FP}+S_{LCO}
\end{equation}
with $S_{LCO}$ given by (\ref{slco}), is $NO$ invariant.\\\\The
Ward identities in the Appendix A can be generalized (see also
\cite{Lemes:2002rc}). As such, the renormalizability is assured,
while the ghost operators still have the same anomalous dimension.
Of course, the relation $\rho=2\zeta$ still holds. Also the
discussion in the previous section can be
repeated\footnote{Although a dynamical gluon mass has up to now
only been calculated for quarkless QCD, the results of
\cite{Verschelde:2001ia,Dudal:2003gu,Dudal:2003vv} could be
generalized to the case with quarks included.}. \\\\ For what
concerns the explicit evaluation of the effective potential in the
Landau gauge, the absence of a counterterm for the ghost operators
(so $Z_{2}=1$) is still valid, just as the relation
$Z_{c}=Z_{g}^{-1}Z_{A}^{-1/2}$. Since the quarks are not
contributing to $\mathcal{W}(\omega,\tau,L)$ at the 1- and 2-loop
level, no new divergences appear at the 1- and 2-loop level, hence
$\delta\rho_{0}$ and $\delta\rho_{1}$ are unchanged in comparison
with the quarkless case. Since \cite{Larin:tp,Gracey:2002yt}
\begin{eqnarray}\label{q4}
    \beta(g^{2})&=& -\varepsilon g^{2}+\left(-\frac{22}{3}N+\frac{4}{3}N_{f}\right)g^{2}\frac{g^{2}}{16\pi^{2}}
    \nonumber\\&+&\left(-\frac{68}{3}N^{2}+\frac{20}{3}N_{f}N+2N_{f}\frac{N^{2}-1}{N}\right)g^{2}\left(\frac{g^{2}}{16\pi^{2}}\right)^{2}+\ldots\nonumber\\
    \gamma_{A}(g^{2})&=&\left(-\frac{13}{6}N+\frac{2}{3}N_{f}\right)\frac{g^{2}}{16\pi^{2}}+
    \left(-\frac{59}{8}N^{2}+\frac{5}{2}N_{f}N+N_{f}\frac{N^{2}-1}{N}\right)\left(\frac{g^{2}}{16\pi^{2}}\right)^{2}+\ldots\nonumber\\
\end{eqnarray}
we now find (again for $N=2$)
\begin{eqnarray}
  \zeta_{0} &=& \frac{3}{2N_{f}-13} \\
  \rho_{0} &=&\frac{6}{2N_{f}-13}\\
  \zeta_{1}&=&\frac{41N_{f}-190}{96(13-2N_{f})\pi^{2}}\\
  \rho_{1}&=&\frac{41N_{f}-190}{48(13-2N_{f})\pi^{2}}
\end{eqnarray}
while the 1-loop effective potential reads
\begin{equation}\label{q5}
    V_{1}(\chi)=\frac{\chi^{2}}{g^{2}\zeta_{0}}\left(1-\frac{\zeta_{1}}{\zeta_{0}}g^{2}\right)
    +\frac{1}{32\pi^{2}}\frac{\chi^{2}}{\zeta_{0}^{2}}\left(\ln\frac{\chi^{2}}{\zeta_{0}^{2}\overline{\mu}^{4}}-3\right)
\end{equation}
with $\chi$ defined as in (\ref{t10}). The minima can be
determined in the same fashion as before, this leads to
\begin{equation}\label{RGE5}
  \left.\frac{g^{2}N}{16\pi^{2}}\right|_{N=2}=\frac{36}{112-29N_{f}}
\end{equation}

\sect{Prospective view on future work} In this section, we would
like to outline some items that deserve further investigation.
\begin{itemize}
\item For simplicity, we have restricted ourselves in this article
to $N=2$. Also, the effective potential has been determined at the
one-loop level, by making use of the $\overline{MS}$ scheme. Then,
as it is apparent from (\ref{RGE5}), the numbers of flavours must
be so that $0\leq N_{f}\leq 3$, in order to have a non-trivial
solution. This can be changed if another renormalization scheme is
chosen. There exist several methods to improve perturbation theory
and minimize the renormalization scheme dependence, for example by
introducing effective charges
\cite{Grunberg:1980ja,Grunberg:1982fw} or by employing the
principle of minimal sensitivity
\cite{Stevenson:1981vj,VanAcoleyen:2001gf}. Also, higher order
computations are in order to improve results. Evidently, 'real
life' QCD will need the generalization to $N=3$.

\item Secondly, we want to comment on the observation that the
ghost condensation gives rise to a tachyonic mass for the gluons
in the Curci-Ferrari gauge \cite{Sawayanagi}. Let us consider this
in more detail in the Landau gauge for $N=2$. The ghost propagator
in the condensed vacuum (\ref{OV})
reads\footnote{$\epsilon^{12}=-\epsilon^{21}=1$, zero otherwise. }
\begin{eqnarray}\label{pros1}
    \left\langle \oc^{a}
    c^{b}\right\rangle_{p}&=&-i\frac{p^{2}\delta^{ab}-\frac{\phi^{3}}{\rho_{0}}\epsilon^{ab}}{p^{4}+\left(\frac{\phi^{3}}{\rho_{0}}\right)^{2}}\hspace{2cm}a,b=1,2\nonumber\\
\left\langle \oc^{3}
    c^{3}\right\rangle_{p}&=&\frac{-i}{p^{2}}
\end{eqnarray}
Following \cite{Sawayanagi}, one can calculate the gauge boson
polarization $\Pi^{ab}_{\mu\nu}$ with this ghost propagator (see
Figure 1), and then one finds an induced tachyonic gluon mass.
Notice that this mass is a loop effect. This observation gave rise
to the conclusion that gluons acquire a tachyonic mass due to the
ghost condensation. It was already recognized in
\cite{Dudal:2002xe} for the Maximal Abelian gauge that the ghost
condensation resulted in a tachyonic mass for the off-diagonal
gluons. In our opinion, this tachyonic mass is more a consequence
of an incomplete treatment than a result \emph{in se}. The gauge
boson polarization was determined with the usual perturbative
gluon propagator (i.e. massless gluons). It was however shown that
gluons get a mass trough a non-vanishing vacuum expectation value
for $\left\langle \frac{1}{2}A^{2}\right\rangle$ in the Landau
gauge \cite{Verschelde:2001ia} or $\left\langle
\frac{1}{2}A^{2}+\alpha \oc c\right\rangle$ in the Curci-Ferrari
gauge \cite{Dudal:2003gu}. The LCO treatment for $\left\langle
\frac{1}{2}A^{2}\right\rangle$ gives a Lagrangian similar to
(\ref{d15}). More precisely, a real \emph{tree level} gluon mass
$m_{\tiny{\textrm{gluon}}}$ is present. It came out that
$m_{\tiny{\textrm{gluon}}}\sim500\textrm{MeV}$
\cite{Verschelde:2001ia}. Therefore, the complete procedure to
analyze the nature of the induced gluon mass should be that of
taking into account the simultaneous presence of both ghost and
gluon condensates, i.e. $\left\langle
f^{abc}\oc^{b}c^{c}\right\rangle$ and
$\left\langle\frac{1}{2}A^{2}\right\rangle$ (or
$\left\langle\frac{1}{2}A^{2}+\alpha\oc c\right\rangle$ in the
Curci-Ferrari gauge). The induced final gluon mass receives
contributions from both condensates, as the gluon propagator gets
modified by the condensate $\left\langle \frac{1}{2}
A^{2}\right\rangle$. The diagram of Figure 1 is thus only part of
the whole set of diagrams contributing to the gluon mass. It is
worth mentioning that a similar mechanism should take place in the
Maximal Abelian gauge
\cite{Dudal:2002xe,Dudal:2002ye,Dudal:2003gu}. In fact, the mixed
gluon-ghost operator $\left\langle\frac{1}{2}A^{2}+\alpha\oc
c\right\rangle$ can be consistently introduced also in this gauge
\cite{Kondo:2001nq,kmsi}. \\\\Summarizing, a complete discussion
of the mass generation for gluons would require a combination the
LCO formalism of this article with that of
\cite{Verschelde:2001ia,Dudal:2003gu} by introducing an extra
source term $\frac{1}{2}KA_{\mu}^{a}A^{\mu a}$ for the operator
$\frac{1}{2}A^{2}$. This will be performed elsewhere, since the
aim of this paper is to discuss the ghost condensates and their
role in the breaking of the $NO$ symmetry.
\begin{figure}[t]\label{fig1}
    \begin{center}
        \scalebox{1}{\includegraphics{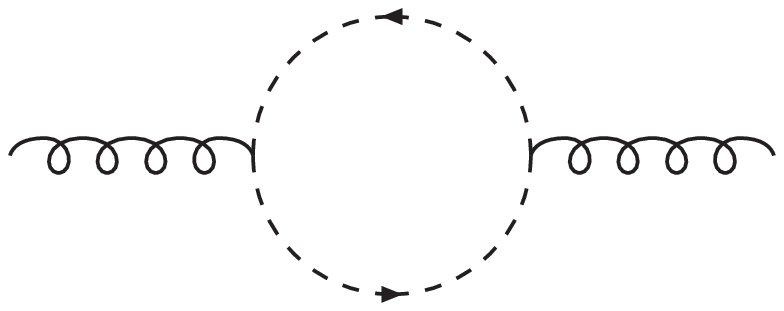}}
        \caption{Diagram relevant for the gauge boson polarization.}
    \end{center}
\end{figure}
\item A third point of interest is the modified infrared behaviour
of the propagators due to the non-vanishing condensates. If one
considers the Landau gauge, the \mbox{Kugo-Ojima} confinement
criterion \cite{Kugo:gm} can be translated into an infrared
enhancement of the ghost propagator, i.e. the ghost propagator
should be more singular than $\frac{1}{p^{2}}$ \cite{Kugo:1995km}.
Recently, much effort has been paid to investigate this criterion
(in the Landau gauge) by means of the Schwinger-Dyson equations,
see e.g.
\cite{Watson:2001yv,Alkofer:2003vj,Lerche:2002ep,Zwanziger:2001kw,Langfeld:2002bg,Kondo:2003sw}
and references therein. Defining the gluon and ghost form factors
from the Euclidean propagators $D_{\mu\nu}(p^{2})$ and $G(p^{2})$
as
\begin{eqnarray}
  D_{\mu\nu}(p^{2}) &=& \left(\delta_{\mu\nu}-\frac{p_{\mu}p_{\nu}}{p^{2}}\right)\frac{Z_{D}(p^{2})}{p^{2}} \nonumber\\
  G(p^{2}) &=& \frac{Z_{G}(p^{2})}{p^{2}}
\end{eqnarray}
it was shown that in the infrared
\begin{eqnarray}
  Z_{D}(p^{2}) &\sim& (p^{2})^{2a} \nonumber\\
  Z_{G}(p^{2}) &\sim& (p^{2})^{-a}
\end{eqnarray}
with $a\approx0.595$
\cite{Alkofer:2003vj,Lerche:2002ep,Zwanziger:2001kw,Langfeld:2002bg}.
As such, the obtained solutions of the Schwinger-Dyson equations
seem to be compatible with the Kugo-Ojima confinement criterion.
Furthermore, these solutions were also in qualitative agreement
with the lattice behaviour (see e.g. \cite{Alkofer:2003vj}). It
would be instructive to investigate to what extent the
Schwinger-Dyson solutions are modified if one would work with the
Landau gauge
action\footnote{$\left\langle\varphi\right\rangle=\frac{g}{2}\left\langle
A_{\mu}^{a}A^{\mu a}\right\rangle$. See
\cite{Verschelde:2001ia,Dudal:2003gu} for the meaning and value of
$\xi$.}
\begin{eqnarray}\label{pros2}
S&=&S_{YM}+S_{GF+FP}+\int d^{4}x\left(-\frac{\varphi^{2}}{2g^{2}\xi} -%
\frac{\sigma ^{a}\overline{\sigma }^{a}}{g^{2}\zeta }-\frac{\phi ^{a}\phi ^{a}}{%
2g^{2}\rho }
\right.  \nonumber   \\
&+&\left.\frac{\varphi}{2g\xi}A_{\mu}^{a}A^{\mu
a}+\frac{\overline{\sigma }^{a}}{2g\zeta}gf^{abc}c^{b}c^{c}
- \frac{\sigma ^{a}}{2g\zeta }gf^{abc}\overline{c}^{b}%
\overline{c}^{c}-\frac{\phi ^{a}}{g\rho }gf^{abc}\overline{c}%
^{b}c^{c}\right.\nonumber\\&-&\left.\frac{1}{8\xi}\left(A_{\mu}^{a}A^{\mu
a}\right)^{2}-\frac{1}{2\rho }g^{2}\left(
f^{abc}\overline{c}^{b}c^{c}\right)
^{2}+\frac{1}{4\zeta }g^{2}f^{abc}c^{b}c^{c}f^{ade}\overline{c}^{d}\overline{c%
}^{e}\right)
\end{eqnarray}
that already incorporates the non-perturbative effects of the
ghost condensates and the gluon condensates, thus also a gluon
mass.\\\\
Very recently, some results for general covariant gauges
concerning the ghost-antighost condensate $\left\langle
c^{a}\overline{c}^{a}\right\rangle$ were presented in
\cite{Alkofer:2003jr} within the Schwinger-Dyson approach. In the
used approximation scheme, it turns out that in case of the linear
gauges, no ghost-antighost condensate seems to exist. It is worth
remarking here that the ghost-antighost condensate $\left\langle
c^{a}\overline{c}^{a}\right\rangle$ is not BRST invariant. It can
be combined with the gluon operator $A^{2}$ to yield the mixed
gluon-ghost dimension two operator
$\frac{1}{2}A^{2}+\alpha\overline{c} c$. To our knowledge, this
operator is on-shell BRST invariant only in the Curci-Ferrari and
in the Maximal Abelian gauge\footnote{In which case the color
index is restricted to the off-diagonal fields.}
\cite{Kondo:2001nq,kmsi,Gripaios:2003xq}. In particular,
concerning the nonlinear Curci-Ferrari gauge, the condensate
$\left\langle\frac{1}{2}A^{2}+\alpha\overline{c} c\right\rangle$
has been proven to show up in the weak coupling
\cite{Dudal:2003gu}. However, no definitive conclusion has been
reached so far about this condensate within the Schwinger-Dyson
framework \cite{Alkofer:2003jr}. Finally, we notice that the ghost
operators $f^{abc}c^{b}c^{c}, f^{abc}\overline{ c}^{b} c^{c}$ and
$f^{abc}\overline{ c}^{b} \overline{ c}^{c}$ we discussed here,
were not considered in \cite{Alkofer:2003jr}. \item We have
discussed the ghost condensation in the Curci-Ferrari gauge.
Originally, the ghost condensates came to attention in the Maximal
Abelian gauge in
\cite{Schaden:1999ew,Schaden:2000fv,Schaden:2001xu,Kondo:2000ey,Dudal:2002xe,Lemes:2002ey,Lemes:2002jv}.
An approach close to the one presented here should be applied to
probe the ghost condensates and their consequences in the Maximal
Abelian gauge too. However, the Maximal Abelian gauge is a bit
more tricky to handle, see e.g. \cite{Dudal:2003gu} for some more
comments on this. \item So far, the gauges where the ghost
condensation takes place, all have the $NO$ symmetry. The
important question rises if the ghost condensation only takes
place in gauges possessing the $NO$ symmetry? In order to do so,
one should first investigate if external sources for the ghost
operators can be introduced without spoiling the
renormalizability. Assuming that the condensation takes place in
gauges without the extra $NO$ invariance, we are however no longer
able to relate the different channels by a $NO$ transformation.
Neither would we be able anymore to define e.g. a 'new'
Faddeev-Popov charge in non-Overhauser like vacua. Therefore, one
might speculate that the enlarged symmetry structure of Yang-Mills
theory is necessary to make sense out of the theory, at least if
the ghost condensation occurs.
\end{itemize}

\sect{Conclusion} In this article, we considered Yang-Mills theory
in the Curci-Ferrari gauge and as a limiting case, in the Landau
gauge. These gauges possess a global continuous symmetry,
generated by the $NO$ algebra. This algebra is built out of the
(anti-)BRST transformation and of the $SL(2,R)$ algebra. By
combining the local composite operator formalism with the
algebraic renormalization technique, we have proven that a ghost
condensation $\grave{\textrm{a}}$ la $\left\langle
f^{abc}c^{b}c^{c}\right\rangle$, $\left\langle
f^{abc}\oc^{b}\oc^{c}\right\rangle$ (BCS channel) and
$\left\langle f^{abc}\oc^{b}c^{c}\right\rangle$ (Overhauser
channel) occurs. It has been shown that different vacua are
possible, with the Overhauser and BCS vacuum as two special
choices. The ghost condensates (partially) break the $NO$
symmetry. We have discussed the BRST and the ghost number symmetry
in the condensed vacua. We paid attention to the global $SU(N)$
color symmetry and to the absence of Goldstone bosons in the
physical spectrum. We also briefly discussed the generalization to
the case when quark fields are included. We ended with some
comments on future research.

\sect{Appendix A}

\subsection{Ward identities for the $NO$ algebra in the Curci-Ferrari gauge}

The renormalizability of the Curci-Ferrari gauge is well
established \cite{Baulieu:sb,Gracey:2002yt,kmsi}. In this Appendix
we show that the introduction of a suitable set of external
sources allows to write down Ward identities for all the
generators of the $NO$\ algebra. In particular, these
Ward identities will imply that all ghost polynomials $f^{abc}c^{b}c^{c}$,  $%
f^{abc}\overline{c}^{b}\overline{c}^{c}$,
$f^{abc}\overline{c}^{b}c^{c}$ have the same anomalous
dimension.\\\\In order to write down the functional identities for
the $NO$ algebra, we
need to introduce three more external sources $\Omega _{\mu }^{a}$, $%
\overline{\Omega }_{\mu }^{a}$, $\vartheta _{\mu }^{a}$ with dimensions $%
\left( 2,2,1\right) $, coupled to the nonlinear BRST\ and
anti-BRST\ variations of the gauge field $A_{\mu }^{a}$.
\begin{equation}
S_{ext}=s\overline{s}\int d^{4}x\left( \vartheta ^{a\mu }A_{\mu }^{a}+\frac{%
\gamma }{2}\vartheta ^{a\mu }\vartheta _{\mu }^{a}\right)
\label{om}
\end{equation}
Notice that the coefficient $\gamma $ is allowed by power
counting, since the term $\vartheta ^{a\mu }\vartheta _{\mu }^{a}$
has dimension $2$.  The
generators of the $NO$ algebra act on $\Omega _{\mu }^{a}$, $\overline{%
\Omega }_{\mu }^{a}$, $\vartheta _{\mu }^{a}$ as
\begin{eqnarray}
s\vartheta _{\mu }^{a} &=&\overline{\Omega }_{\mu }^{a}  \label{sa} \\
s\overline{\Omega }_{\mu }^{a} &=&s\Omega _{\mu }^{a}=0 \nonumber
\end{eqnarray}

\begin{eqnarray}
\overline{s}\vartheta _{\mu }^{a} &=&-\Omega _{\mu }^{a}  \label{asa} \\
\overline{s}\Omega _{\mu }^{a} &=&\overline{s}\overline{\Omega
}_{\mu }^{a}=0  \nonumber
\end{eqnarray}
\begin{eqnarray}
\delta \Omega _{\mu }^{a} &=&-\overline{\Omega }_{\mu }^{a}  \label{da} \\
\delta \vartheta _{\mu }^{a} &=&\delta \Omega _{\mu }^{a}=0
\nonumber
\end{eqnarray}
and
\begin{eqnarray}
\overline{\delta }\overline{\Omega }_{\mu }^{a} &=&-\Omega _{\mu
}^{a}
\label{ada} \\
\overline{\delta }\vartheta _{\mu }^{a} &=&\overline{\delta
}\Omega _{\mu }^{a}=0  \nonumber
\end{eqnarray}
Therefore, for $S_{ext}$ one gets

\begin{equation}
S_{ext}=\int d^{4}x\left( -\Omega ^{\mu a}D_{\mu }^{ab}c^{b}-\overline{%
\Omega }^{\mu a}D_{\mu }^{ab}\overline{c}^{b}+\gamma \Omega ^{\mu a}%
\overline{\Omega }_{\mu }^{a}-\vartheta ^{a\mu }D_{\mu
}^{ab}b^{b}+gf^{abc}\vartheta ^{a\mu }\left( D_{\mu
}^{bd}c^{d}\right) \overline{c}^{c}\right)   \label{gamma}
\end{equation}
>From this expression, it can be seen that the parameter $\gamma $
is needed to account for the behavior of the two-point Green
function $\left\langle
\left( D_{\mu }^{ab}c^{b}(x)\right) \left( D_{\nu }^{cd}\overline{c}%
^{d}(y)\right) \right\rangle $, which is deeply related to the
Kugo-Ojima criterion. In other words, the coefficient $\gamma $ is
the LCO parameter for this Green function. \\\\We can now
translate the whole $NO$ algebra into functional identities, which
will be the starting point for the algebraic characterization of
the allowed counterterm. It turns out thus that, in the
Curci-Ferrari gauge, the
complete action $\Sigma $%
\begin{eqnarray}
\Sigma  &=&S_{YM}+S_{GF+FP}+S_{LCO}+S_{ext}  \nonumber \\
&=&-\frac{1}{4}\int d^{4}xF_{\mu \nu }^{a}F^{a\mu \nu
}+s\overline{s}\int
d^{4}x\left(\frac{}{} \lambda ^{a}c^{a}+\zeta \lambda ^{a}\eta ^{a}+\eta ^{a}%
\overline{c}^{a}\right.   \nonumber \\
&&\left. \,\,\,\,\,\,\,\,\,\,\,\,\,\,\,\,\,\,\,\,\,\,\,\,-\frac{\alpha }{2}%
c^{a}\overline{c}^{a}+\vartheta ^{a\mu }A_{\mu
}^{a}+\frac{1}{2}A_{\mu }^{a}A^{a\mu }+\frac{\gamma }{2}\vartheta
^{a\mu }\vartheta _{\mu }^{a}\right)   \label{cacf}
\end{eqnarray}
is constrained by the following identities:

\begin{itemize}
\item  the Slavnov-Taylor identity
\begin{equation}
\mathcal{S}(\Sigma )=0  \label{stl}
\end{equation}
\begin{eqnarray}
\mathcal{S}(\Sigma ) &=&\int d^{4}x\left( \left( \frac{\delta \Sigma }{%
\delta \Omega ^{a\mu }}-\gamma \overline{\Omega }_{\mu }^{a}\right) \frac{%
\delta \Sigma }{\delta A_{\mu }^{a}}+\left( \frac{\delta \Sigma
}{\delta L^{a}}-\zeta \tau ^{a}\right) \frac{\delta \Sigma
}{\delta c^{a}}\right.
\nonumber \\
&&\;\;\;\;\;\;\;\;\;\left. +b^{a}\frac{\delta \Sigma }{\delta \overline{c}%
^{a}}+\tau ^{a}\frac{\delta \Sigma }{\delta \eta ^{a}}+\omega ^{a}\frac{%
\delta \Sigma }{\delta \lambda ^{a}}+\overline{\Omega }_{\mu }^{a}\frac{%
\delta \Sigma }{\delta \vartheta _{\mu }^{a}}\right) \label{stcfe}
\end{eqnarray}

\item  the anti-Slavnov-Taylor identity
\begin{equation}
\overline{\mathcal{S}}(\Sigma )=0  \label{ast}
\end{equation}
\begin{eqnarray}
\overline{\mathcal{S}}(\Sigma ) &=&\int d^{4}x\left( \left(
\frac{\delta \Sigma }{\delta \overline{\Omega }^{a\mu }}+\gamma
\Omega _{\mu }^{a}\right)
\frac{\delta \Sigma }{\delta A_{\mu }^{a}}-\left( \frac{\delta \Sigma }{%
\delta \tau ^{a}}-\zeta L^{a}\right) \frac{\delta \Sigma }{\delta \overline{c%
}^{a}}\right. \;  \nonumber \\
&&\;\;\;\;\;\left. -\left( b^{a}+\frac{\delta \Sigma }{\delta \omega ^{a}}%
-2\zeta \omega ^{a}\right) \frac{\delta \Sigma }{\delta
c^{a}}-\frac{\delta
\Sigma }{\delta \eta ^{a}}\frac{\delta \Sigma }{\delta b^{a}}+L^{a}\frac{%
\delta \Sigma }{\delta \lambda ^{a}}\right.   \nonumber \\
&&\;\;\;\;\;\left. -\omega ^{a}\frac{\delta \Sigma }{\delta \eta ^{a}}%
-\Omega ^{a\mu }\frac{\delta \Sigma }{\delta \vartheta ^{a\mu
}}\right) \label{astc}
\end{eqnarray}

\item  the $\delta $ Ward identity
\begin{equation}
\mathcal{W}(\Sigma )=0  \label{dwl}
\end{equation}
with
\begin{eqnarray}
\mathcal{W}(\Sigma ) &=&\int d^{4}x\left( c^{a}\frac{\delta \Sigma
}{\delta \overline{c}^{a}}+\left( \frac{\delta \Sigma }{\delta
L^{a}}-\zeta \tau ^{a}\right) \frac{\delta \Sigma }{\delta
b^{a}}+2\omega ^{a}\frac{\delta
\Sigma }{\delta L^{a}}\right. \;  \nonumber \\
&&\;\;\;\;\;\;\;\;\left. -\tau ^{a}\frac{\delta \Sigma }{\delta \omega ^{a}}%
-\eta ^{a}\frac{\delta \Sigma }{\delta \lambda
^{a}}-\overline{\Omega }_{\mu }^{a}\frac{\delta \Sigma }{\delta
\Omega _{\mu }^{a}}\right)   \label{di}
\end{eqnarray}

\item  the $\overline{\delta }$ Ward identity
\begin{equation}
\overline{\mathcal{W}}(\Sigma )=0  \label{adid}
\end{equation}
\begin{eqnarray}
\overline{\mathcal{W}}(\Sigma ) &=&\int d^{4}x\left( \overline{c}^{a}\frac{%
\delta \Sigma }{\delta c^{a}}-\left( \frac{\delta \Sigma }{\delta \tau ^{a}}%
-\zeta L^{a}\right) \frac{\delta \Sigma }{\delta b^{a}}-2\omega ^{a}\frac{%
\delta \Sigma }{\delta \tau ^{a}}\right. \;  \nonumber \\
&&\;\;\;\;\;\;\;\;\left. +L^{a}\frac{\delta \Sigma }{\delta \omega ^{a}}%
-\lambda ^{a}\frac{\delta \Sigma }{\delta \eta ^{a}}-\Omega _{\mu }^{a}\frac{%
\delta \Sigma }{\delta \overline{\Omega }_{\mu }^{a}}\right)
\label{adop}
\end{eqnarray}
\end{itemize}

\subsection{Algebraic characterization of the invariant counterterm in the
Curci-Ferrari gauge}

The most general local invariant counterterm compatible with both
Slavnov-Taylor and anti-Slavnov-Taylor identities $\left( \mathrm{{\ref{stl}}%
}\right) $, $\left( \mathrm{{\ref{ast}}}\right) $ can be written
as
\begin{eqnarray}
\Sigma ^{c} &=&-\frac{\sigma }{4}\int d^{4}xF_{\mu \nu }^{a}F^{a\mu \nu }+%
\mathcal{B}\overline{\mathcal{B}}\int d^{4}x\left( \frac{}{}
a_{1}\lambda ^{a}c^{a}+a_{2}\eta ^{a}\lambda ^{a}+a_{3}\eta
^{a}\overline{c}^{a}\right.
\nonumber \\
&&\left. \,\,\,\,\,\,\,\,\,\,\,\,\,\,\,\,\,\,\,\,\,\,\,\,+\frac{a_{4}}{2}%
c^{a}\overline{c}^{a}+a_{5}\vartheta _{\mu }^{a}A^{a\mu }+\frac{a_{6}}{2}%
A^{a\mu }A_{\mu }^{a}+\frac{a_{7}}{2}\vartheta _{\mu
}^{a}\vartheta ^{a\mu }\right)   \label{invc}
\end{eqnarray}
where $\sigma $, $a_{1}$, $a_{2}$, $a_{3}$, $a_{4}$, $a_{5}$, $a_{6}$, $%
a_{7}$ are free parameters and $\mathcal{B}$,
$\overline{\mathcal{B}}$ denote the linearized nilpotent operators
\begin{eqnarray}
\mathcal{B} &=&\int d^{4}x\left( \frac{\delta \Sigma }{\delta A_{\mu }^{a}}%
\frac{\delta }{\delta \Omega ^{a\mu }}+\left( \frac{\delta \Sigma
}{\delta
\Omega ^{a\mu }}-\gamma \overline{\Omega }_{\mu }^{a}\right) \frac{\delta }{%
\delta A_{\mu }^{a}}+\left( \frac{\delta \Sigma }{\delta
L^{a}}-\zeta \tau
^{a}\right) \frac{\delta }{\delta c^{a}}\right.   \nonumber \\
&&\;\;\;\;\;\;\;\;\;\;\left. +\frac{\delta \Sigma }{\delta c^{a}}\frac{%
\delta }{\delta L^{a}}+b^{a}\frac{\delta }{\delta \overline{c}^{a}}+\tau ^{a}%
\frac{\delta }{\delta \eta ^{a}}+\omega ^{a}\frac{\delta }{\delta
\lambda ^{a}}+\overline{\Omega }_{\mu }^{a}\frac{\delta }{\delta
\vartheta _{\mu }^{a}}\right)   \label{bb}
\end{eqnarray}
and

\begin{eqnarray}
\overline{\mathcal{B}} &=&\int d^{4}x\left( \frac{\delta \Sigma
}{\delta A_{\mu }^{a}}\frac{\delta }{\delta \overline{\Omega
}_{\mu }^{a}}+\left( \frac{\delta \Sigma }{\delta \overline{\Omega
}^{a\mu }}+\gamma \Omega _{\mu }^{a}\right) \frac{\delta }{\delta
A_{\mu }^{a}}-\left( \frac{\delta \Sigma
}{\delta \tau ^{a}}-\zeta L^{a}\right) \frac{\delta }{\delta \overline{c}^{a}%
}-\frac{\delta \Sigma }{\delta \overline{c}^{a}}\frac{\delta
}{\delta \tau
^{a}}\right.   \nonumber \\
&&\left. -\left( b^{a}+\frac{\delta \Sigma }{\delta \omega
^{a}}-2\zeta \omega ^{a}\right) \frac{\delta }{\delta
c^{a}}-\frac{\delta \Sigma }{\delta c^{a}}\frac{\delta }{\delta
\omega ^{a}}-\frac{\delta \Sigma }{\delta \eta
^{a}}\frac{\delta }{\delta b^{a}}-\frac{\delta \Sigma }{\delta b^{a}}\frac{%
\delta }{\delta \eta ^{a}}\right.   \nonumber \\
&&\left. +L^{a}\frac{\delta }{\delta \lambda ^{a}}-\omega ^{a}\frac{\delta }{%
\delta \eta ^{a}}-\Omega ^{a\mu }\frac{\delta }{\delta \vartheta ^{a\mu }}%
\right)   \label{abb}
\end{eqnarray}
>From the $\delta $ and $\overline{\delta }$ Ward identities $\left( \mathrm{{%
\ref{dwl}}}\right) $, $\left( \mathrm{{\ref{adid}}}\right) $ it
follows that
\begin{equation}
a_{3}=a_{1}  \label{a3a1}
\end{equation}
so that the final expression for $\left(
\mathrm{{\ref{invc}}}\right) $ becomes
\begin{eqnarray}
\Sigma ^{c} &=&-\frac{\sigma }{4}\int d^{4}xF_{\mu \nu }^{a}F^{a\mu \nu }+%
\mathcal{B}\overline{\mathcal{B}}\int d^{4}x\left(\frac{}{}
a_{1}\lambda ^{a}c^{a}+a_{2}\eta ^{a}\lambda ^{a}+a_{1}\eta
^{a}\overline{c}^{a}\right.
\nonumber \\
&&\left. \,\,\,\,\,\,\,\,\,\,\,\,\,\,\,\,\,\,\,\,\,\,\,\,\,+\frac{a_{4}}{2}%
c^{a}\overline{c}^{a}+a_{5}\vartheta _{\mu }^{a}A^{a\mu }+\frac{a_{6}}{2}%
A^{a\mu }A_{\mu }^{a}+\frac{a_{7}}{2}\vartheta _{\mu
}^{a}\vartheta ^{a\mu }\right)   \label{finvc}
\end{eqnarray}
The coefficients  $\sigma $, $a_{1}$, $a_{2}$, $a_{4}$, $a_{5}$, $a_{6}$, $%
a_{7}$ are easily seen to correspond to a multiplicative
renormalization of
the coupling constant $g$, of the gauge and LCO\ parameters $\alpha $, $%
\zeta $, $\gamma $, of the fields and external sources.  In
particular, the coefficients $\sigma $ and $a_{5}$ are related to
the renormalization of the gauge coupling constant $g$ and of the
gauge field $A_{\mu }^{a}$, as it is apparent from
\begin{equation}
\mathcal{B}\overline{\mathcal{B}}\int d^{4}x\vartheta _{\mu }^{a}A^{a\mu }=-%
\mathcal{N}_{A}\Sigma \,\,\,  \label{a5}
\end{equation}
where $\mathcal{N}_{A}$ stands for the invariant counting operator

\begin{equation}
\mathcal{N}_{A}=\int d^{4}x\left( A_{\mu }^{a}\frac{\delta
}{\delta A_{\mu
}^{a}}-\Omega _{\mu }^{a}\frac{\delta }{\delta \Omega _{\mu }^{a}}-\overline{%
\Omega }_{\mu }^{a}\frac{\delta }{\delta \overline{\Omega }_{\mu }^{a}}%
-\vartheta _{\mu }^{a}\frac{\delta }{\delta \vartheta _{\mu
}^{a}}\right) +\gamma \frac{\partial }{\partial \gamma }\,\,\,
\label{na}
\end{equation}
The coefficient $a_{4}$ corresponds to the renormalization of the
gauge parameter $\alpha $, indeed

\begin{equation}
\mathcal{B}\overline{\mathcal{B}}\int d^{4}x\frac{1}{2}c^{a}\overline{c}%
^{a}=-\frac{\partial \Sigma }{\partial \alpha }  \label{a4}
\end{equation}
The coefficient $a_{2}$ is associated to the renormalization of
the LCO parameter $\zeta ,$ which follows from
\begin{equation}
\mathcal{B}\overline{\mathcal{B}}\int d^{4}x\eta ^{a}\lambda ^{a}=-\mathcal{N%
}_{\zeta }\Sigma   \label{a22}
\end{equation}
with
\begin{equation}
\mathcal{N}_{\zeta }=\zeta \frac{\partial }{\partial \zeta }+\int
d^{4}x\left( \omega ^{a}\frac{\delta }{\delta b^{a}}-\eta ^{a}\frac{\delta }{%
\delta c^{a}}+\lambda ^{a}\frac{\delta }{\delta
\overline{c}^{a}}\right)  \label{nxi}
\end{equation}
The coefficient $a_{1}$ is related to the\thinspace anomalous
dimensions of all ghost operators, namely

\begin{equation}
\mathcal{B}\overline{\mathcal{B}}\int d^{4}x\left( \lambda
^{a}c^{a}+\eta ^{a}\overline{c}^{a}\right) =\mathcal{N}_{L}\Sigma
  \label{a111}
\end{equation}
where
\begin{eqnarray}
\mathcal{N}_{L} &=&\;\int d^{4}x\left( L^{a}\frac{\delta }{\delta L^{a}}%
+\tau ^{a}\frac{\delta }{\delta \tau ^{a}}+\lambda
^{a}\frac{\delta }{\delta
\lambda ^{a}}+\omega ^{a}\frac{\delta }{\delta \omega ^{a}}+\eta ^{a}\frac{%
\delta }{\delta \eta ^{a}}\right.   \nonumber \\
&&\left. \,\,\,\,\,\,\,\,\,\,\,\,\,\,\,\,\,\,\,\,\,-b^{a}\frac{\delta }{%
\delta b^{a}}-\overline{c}^{a}\frac{\delta }{\delta \overline{c}^{a}}-c^{a}%
\frac{\delta }{\delta c^{a}}\right) -2\zeta \frac{\partial }{\partial \zeta }%
\,\,\,  \label{nl}
\end{eqnarray}
The renormalization of the LCO parameter $\gamma $ is given by the
coefficient $a_{7}$, as can be seen from
\begin{equation}
\mathcal{B}\overline{\mathcal{B}}\int d^{4}x\left(
\frac{1}{2}\vartheta _{\mu }^{a}\vartheta ^{a\mu }\right) =\left(
\frac{\partial }{\partial
\gamma }-\int d^{4}x\,\vartheta _{\mu }^{a}\frac{\delta }{\delta A_{\mu }^{a}%
}\right) \Sigma \,\,\,  \label{ngam}
\end{equation}
Finally, the anomalous dimension of the ghost $c^{a}\,$and the antighost $%
\overline{c}^{a}$ are obtained from the coefficient $a_{6}$

\begin{equation}
\mathcal{B}\overline{\mathcal{B}}\int d^{4}x\left(
\frac{1}{2}A^{a\mu }A_{\mu }^{a}\right) =\mathcal{N}_{c}\Sigma
\,\,\,  \label{rghost}
\end{equation}
with

\begin{eqnarray}
\mathcal{N}_{c} &=&\;\int d^{4}x\left( \frac{1}{2}\,c^{a}\frac{\delta }{%
\delta c^{a}}+\frac{1}{2}\,\overline{c}^{a}\frac{\delta }{\delta \overline{c}%
^{a}}+b^{a}\frac{\delta }{\delta b^{a}}-L^{a}\frac{\delta }{\delta L^{a}}%
-\tau ^{a}\frac{\delta }{\delta \tau ^{a}}-\omega ^{a}\frac{\delta
}{\delta
\omega ^{a}}\right.   \nonumber \\
&&\left. \,\,\,\,\,\,\,\,\,\,\,\,\,\,\,\,\,\,\,\,\,-\frac{3}{2}\lambda ^{a}%
\frac{\delta }{\delta \lambda ^{a}}-\frac{3}{2}\eta ^{a}\frac{\delta }{%
\delta \eta ^{a}}\right) -2\alpha \frac{\partial }{\partial \alpha
}+2\zeta \frac{\partial }{\partial \zeta }\,\,\,  \label{ncc}
\end{eqnarray}
>From expressions $\left( \mathrm{{\ref{nl}}}\right) $ and $%
\left( \mathrm{{\ref{ncc}}}\right) $ one sees that all sources
$L^{a},\tau ^{a},$ and $\omega ^{a}$ renormalize in the same way,
which means that all
composite ghost polynomials $f^{abc}c^{b}c^{c}$, $f^{abc}\overline{c}^{b}%
\overline{c}^{c}$, $f^{abc}\overline{c}^{b}c^{c}$  have indeed the
same
anomalous dimension. This result is a consequence of the relationship $%
\left( \mathrm{{\ref{a3a1}}}\right) $ which, of course, stems from
the existence of the $NO$ algebra.

\sect{Appendix B} In order to construct the 1-loop effective
potential, we need the values of $\zeta_{0}$, $\rho_{0}$,
$\zeta_{1}$ and $\rho_{1}$. These can be calculated as soon we
know the divergences proportional to $\omega^{2}$ and $L\tau$ when
the generating functional corresponding to the action (\ref{d1})
is calculated. In principle, it is sufficient to calculate the
divergences proportional to $\omega^{2}$ since the $NO$ invariance
leads to $\rho=2\zeta$. Therefore, we can restrict ourselves to
the diagrams with only the source $\omega$ connected. Let us write
\begin{equation}\label{ttt}
    \delta\rho=\delta\rho_{0}g^{2}+\delta\rho_{1}g^{4}+\cdots
\end{equation}
For $N=2$ and $\alpha=0$, the ghost propagator reads
\begin{equation}\label{b1}
    \left\langle \oc^{a}c^{b}\right\rangle_{p}=i\frac{-p^{4}\delta^{ab}+p^{2}g\epsilon^{abc}\omega^{c}-g^{2}\omega^{a}\omega^{b}}{p^{2}\left(p^{4}+g^{2}\omega^{2}\right)}
\end{equation}
while the gluon propagator is given by
\begin{equation}\label{b1bis}
    \left\langle
    A_{\mu}^{a}A_{\nu}^{b}\right\rangle_{p}=\frac{-i\delta^{ab}}{p^{2}}\left(g_{\mu\nu}-\frac{p_{\mu}p_{\nu}}{p^{2}}\right)
\end{equation}
The ghost-antighost-gluon vertex equals
\begin{equation}\label{b1tris}
    g\epsilon^{abc}p_{\mu}
\end{equation}
The relevant vacuum bubbles\footnote{The diagrams containing a
counterterm are not shown.} are shown in Figure 2.
        \begin{figure}[t]\label{fig2}
    \begin{center}
        \scalebox{1}{\includegraphics{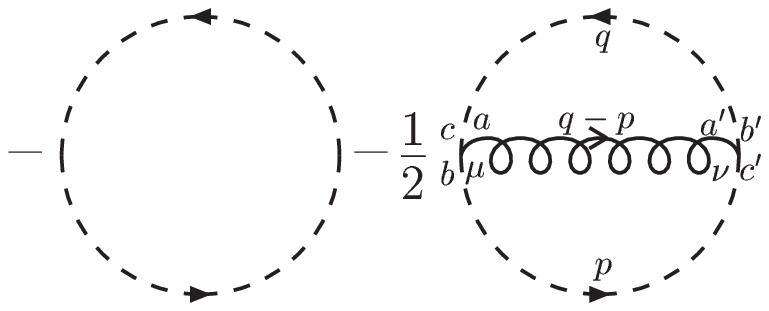}}
        \caption{Vacuum bubbles up to 2-loop order, giving divergences proportional to $\omega^{2}$.}
    \end{center}
    \end{figure}
At 1-loop, we find a contribution to $\mathcal{W}(\omega,\tau,L)$,
given by
\begin{equation}\label{b2}
    -i\int \frac{d^{d}p}{(2\pi)^{d}}\ln\left(p^{4}+g^{2}\omega^{2}\right)
\end{equation}
Performing a Wick rotation\footnote{If one would like to avoid a
Wick rotation, one could have started immediately from the
Euclidean Yang-Mills action.} and employing the $\overline{MS}$
scheme, this leads to a divergence given by
\begin{equation}\label{b3}
    g^{2}\omega^{2}\frac{1}{32\pi^{2}}\frac{4}{\varepsilon}
\end{equation}
Hence
\begin{equation}\label{b4}
    \delta\rho_{0}=-\frac{1}{4\pi^{2}}\frac{1}{\varepsilon}
\end{equation}
At 2-loops, the contribution to $\mathcal{W}(\omega,\tau,L)$ is
obtained by computing the second diagram of Figure 2, yielding
\begin{eqnarray}\label{b5}
    \mathcal{I}=\frac{1}{2}ig^{2}\epsilon^{abc}\epsilon^{a'b'c'}\int
    \frac{d^{d}p}{(2\pi)^{d}}\frac{d^{d}q}{(2\pi)^{d}}\left[p_{\mu}q_{\nu}
    \frac{-i\delta^{aa'}}{(p-q)^{2}}\left(g_{\mu\nu}-\frac{(p-q)_{\mu}(p-q)_{\nu}}{(p-q)^{2}}\right)\right.\nonumber\\
    \left.\times i\frac{-p^{4}\delta^{bc'}+gp^{2}\epsilon^{bc'e}\omega^{e}-g^{2}\omega^{b}\omega^{c'}}{p^{2}(p^{4}+g^{2}\omega^{2})}
    \times
    i\frac{-q^{4}\delta^{b'c}+gq^{2}\epsilon^{b'ce}\omega^{e}-g^{2}\omega^{b'}\omega^{c}}{q^{2}(q^{4}+g^{2}\omega^{2})}\right]
\end{eqnarray}
Working out the color algebra, one finds
\begin{eqnarray}\label{2x}
    \mathcal{I}&=&-\frac{g^{2}}{2}\int
    \frac{d^{d}p}{(2\pi)^{d}}\frac{d^{d}q}{(2\pi)^{d}}\left[
    \frac{p_{\mu}q_{\nu}}{(p-q)^{2}}\left(g_{\mu\nu}-\frac{(p-q)_{\mu}(p-q)_{\nu}}{(p-q)^{2}}\right)
    \right.\nonumber\\&\times&\left.\frac{-6p^{4}q^{4}-2g^{2}\omega^{2}(p^{4}+q^{4}-p^{2}q^{2})}{p^{2}q^{2}(p^{4}+g^{2}
    \omega^{2})(q^{4}+g^{2}\omega^{2})}\right]
\end{eqnarray}
This integral $\mathcal{I}$ has been calculated in two steps:
first all tensor integrals have been reduced to a combination of
scalar master integrals, applying simple algebraic rearrangements
of the scalar products which appear in the numerator of the
integrand; all master integrals are vacuum integrals, i.e. with
vanishing external momentum; they have been replaced by their
explicit expression in terms of special functions
\cite{Davydychev:1992mt} and expanded in powers of $\varepsilon$.
The calculation has been done with the \verb"Mathematica" packages
\emph{DiagExpand} and \emph{ProcessDiagram}. We find
\begin{equation}\label{mi}
    \mathcal{I}=\frac{g^{4}\omega^{2}}{(16\pi^{2})^{2}}\left(\frac{6}{\varepsilon^{2}}+\frac{17}{2\varepsilon}-\frac{6}{\varepsilon}\ln\frac{g\omega}{\overline{\mu}^{2}}+\textrm{finite}\right)
\end{equation}
We also have to take the counterterm information into
account\footnote{The corresponding diagram looks like the first
one of Figure 2.}. Since there is no counterterm $\propto
\omega^{a} gf^{abc}\overline{c}^{b}c^{c}$ in the Landau gauge, the
only counterterm that will contribute at the order we are working,
is
\begin{equation}\label{aa}
    \delta Z_{c}\overline{c}^{a}\partial_{\mu}\partial^{\mu}\delta^{ab}c^{b}
\end{equation}
where \cite{Gracey:2002yt}
\begin{equation}\label{ren}
    \delta Z_{c}=\frac{3}{2}\frac{g^{2}N}{16\pi^{2}}\frac{1}{\varepsilon}+\cdots\equiv z_{c}^{1}g^{2}+\cdots
\end{equation}
This leads to a contribution
\begin{equation}\label{b2}
    \left(-2z_{c}^{1}g^{2}\right)\left(-i\int
    \frac{d^{d}p}{(2\pi)^{d}}\ln\left(p^{4}+g^{2}\omega^{2}\right)\right)
\end{equation}
Or
\begin{equation}\label{tata}
    \left[-2z_{c}^{1}g^{2}\right]\left[-\frac{g^{2}\omega^{2}}{32\pi^{2}}\left(-\frac{4}{\varepsilon}+2\ln\frac{g\omega}{\overline{\mu}^{2}}-3\right)\right]
=\frac{g^{4}\omega^{2}}{(16\pi^{2})^{2}}\left(-\frac{12}{\varepsilon^{2}}-\frac{9}{\varepsilon}+\frac{6}{\varepsilon}\ln\frac{g\omega}{\overline{\mu}^{2}}+\textrm{finite}\right)
\end{equation}
Hence, the complete 2-loop contribution to
$\mathcal{W}(\omega,\tau,L)$ yields
\begin{equation}\label{fin}
    (\ref{mi})+(\ref{tata})=\frac{g^{4}\omega^{2}}{(16\pi^{2})^{2}}\left(-\frac{6}{\varepsilon^{2}}-\frac{1}{2\varepsilon}+\textrm{finite}\right)
\end{equation}
A good internal check of the calculations is that the terms
proportional to
$\frac{1}{\varepsilon}\ln\frac{g\omega}{\overline{\mu}^{2}}$ are
cancelled. Finally, we find that
\begin{equation}\label{b4}
    \delta\rho_{1}=\frac{1}{\left(16\pi^{2}\right)^{2}}\left(\frac{1}{\varepsilon}+\frac{12}{\varepsilon^{2}}\right)
\end{equation}

\section*{Acknowledgments}
D.~D. would like to thank K.~Van Acoleyen for useful discussions.
It is a pleasure for M.~P. to thank R.~Ferrari for enlightening
conversations while this paper was prepared. The Conselho Nacional
de Desenvolvimento Cient\'{i}fico e Tecnol\'{o}gico CNPq-Brazil,
the Funda{\c{c}}{\~{a}}o de Amparo a Pesquisa do Estado do Rio de
Janeiro (Faperj), the SR2-UERJ and the Ministero dell'Istruzione
dell'Universit\'a e della Ricerca - Italy are acknowledged for the
financial support.

\end{document}